\documentclass[12pt,draftclsnofoot,onecolumn]{IEEEtran}
\pdfoutput=1
\usepackage{amsfonts}
\usepackage{CJK}
\usepackage{cite}
\usepackage{graphicx}
\usepackage{caption}
\usepackage{epstopdf}
\usepackage{subfigure}
\usepackage[cmex10]{amsmath}
\usepackage[amsmath,thmmarks]{ntheorem}
\usepackage{amssymb}
\usepackage{ntheorem}
\usepackage{mathabx}
\usepackage{algorithm,float}
\usepackage{algorithmic}
\usepackage{lipsum}
\usepackage{booktabs}
\usepackage{threeparttable}
\usepackage{multicol}
\usepackage{multirow}
\usepackage{xcolor}
\usepackage{bm}
\usepackage{mathabx}
\usepackage{stmaryrd}
\usepackage{upgreek}
\usepackage{url}
\usepackage{wasysym}
\usepackage[labelsep=none]{caption}

\ifCLASSINFOpdf

\else

\fi

\begin{document}

\title{Joint Transmit Precoding and Reflect Beamforming Design for IRS-Assisted MIMO Cognitive Radio Systems}

\author{Weiheng~Jiang,~\IEEEmembership{Member,~IEEE,}
        Yu~Zhang,
        Jun~Zhao,~\IEEEmembership{Member,~IEEE,}
        Zehui~Xiong,~\IEEEmembership{Member,~IEEE,}
        Zhiguo Ding,~\IEEEmembership{Fellow,~IEEE,} 
\thanks{Weiheng Jiang and Yu Zhang are with the School of Microelectronics and Communication Engineering, Chongqing University, Chongqing, China, (email: whjiang@cqu.edu.cn, zhangyu2014@cqu.edu.cn). J. Zhao is with the School of Computer Science and Engineering, Nanyang Technological University, Singapore (e-mail: junzhao@ntu.edu.sg). Z. Xiong is with the Pillar of Information Systems Technology and Design, Singapore University of Technology and Design, Singapore (e-mail: zehui\_xiong@sutd.edu.sg). Z. Ding is with the School of Electrical and Electronic Engineering, University of Manchester, Manchester, UK (email: zhiguo.ding@manchester.ac.uk).}}


\maketitle

\begin{abstract}
Cognitive radio (CR) is an effective solution to improve the spectral efficiency (SE) of wireless communications by allowing the secondary users (SUs) to share spectrum with primary users (PUs). Meanwhile, intelligent reflecting surface (IRS) has been recently proposed as a promising approach to enhance SE and energy efficiency (EE) of wireless communication systems through intelligently reconfiguring the channel environment. In this paper, we consider an IRS-assisted downlink CR system, in which a secondary access point (SAP) communicates with multiple SUs without affecting multiple PUs in the primary network and all nodes are equipped with multiple antennas. Our design objective is to maximize the achievable weighted sum rate (WSR) of SUs subject to the total transmit power constraint at the SAP and the interference constraints at PUs, by jointly optimizing the transmit precoding at the SAP and the reflecting coefficients at the IRS. To deal with the complex objective function, the problem is reformulated by employing the well-known weighted minimum mean-square error (WMMSE) method and an alternating optimization (AO)-based algorithm is proposed. Furthermore, a special scenario with only one PU is considered and AO algorithm is adopted again. It is worth mentioning that the proposed algorithm has a much lower computational complexity than the above algorithm without the performance loss. Finally, some numerical simulations have been provided to demonstrate that the proposed algorithm outperforms other benchmark schemes.
\end{abstract}

\begin{IEEEkeywords}
Intelligent Reflecting Surface (IRS), Multiple-Input Multiple-Output (MIMO), Cognitive radio (CR), Resource Allocation, Alternating Optimization (AO).
\end{IEEEkeywords}

\IEEEpeerreviewmaketitle

\section{Introduction}\label{section_1}

\IEEEPARstart{I}{t} is known that the spectral efficiency (SE) and energy efficiency (EE) are the two essential criteria for designing future wireless networks \cite{DBLP:journals/jsac/JoungHS14}. Cognitive radio (CR) has been proposed as one effective way to enhance the radio SE and EE \cite{DBLP:journals/wc/MitolaM99}. Moreover, it has great potential in reducing the cost and the complexity and as well as energy consumption of the future 5G technologies such as massive MIMO with excessive antennas, and also can support the development of the sustainable and green wireless networks in the coming years \cite{DBLP:journals/cm/LiaskosNTPIA18}. Meanwhile, recently, a new technology following the development of the Micro-Electro-Mechanical Systems (MEMS) named as intelligent reflecting surfaces (IRS) has been proposed, which can be reconfigured to achieve a smart wireless propagation environment via software-controlled reflection \cite{2016A, DBLP:journals/ejwcn/RenzoDHZAYSAHGR19, DBLP:journals/cm/WuZ20}.

On the one hand, for the IRS assisted wireless communications, in \cite{DBLP:conf/globecom/WuZ18}, the problem of jointly optimizing the access point (AP) active beamforming and IRS passive beamforming with AP transmission power constraint to maximize the received signal power for one pair of transceivers was discussed. Based on the semidefinite relaxation (SDR) and the alternate optimization (AO), both the centralized algorithm and distributed algorithm were proposed therein. The work \cite{DBLP:journals/twc/WuZ19} extended the previous work to the multi-users scenario with the individual signal-to-noise ratio (SNR) constraints, where the joint optimization of the AP active beamforming and IRS passive beamforming was developed to minimize the total AP transmission power, and two suboptimal algorithms with different performance-complexity tradeoff were presented. Huang et al. \cite{DBLP:conf/icassp/HuangZDY18} considered the IRS-based multiple-input single-output (MISO) downlink multi-user communications for an outdoor environment, where the base station (BS) transmission power and IRS phase shift were optimized with BS transmission power constraint and user signal-to-interference-and-noise-ratio (SINR) constraint to maximize the sum system rate. Since the formulated resource allocation problem is non-convex,  Majorization-Minimization (MM) and AO were jointly applied, and the convergence of the proposed algorithm was analyzed. Different from the continuous phase shift assumption of the IRS reflecting elements in existing studies, \cite{DBLP:journals/tcom/WuZ20} considered that each IRS reflecting element can only achieve discrete phase shift and the joint optimization of the multi-antenna AP beamforming and IRS discrete phase shift was discussed under the same scenario as \cite{DBLP:conf/globecom/WuZ18}. Then the performance loss caused by the IRS discrete phase shift was quantitatively analyzed via comparing with the IRS continuous phase shift. It is surprising that, the results have shown that as the number of IRS reflecting elements approaches infinity, the system can obtain the same square power gain as IRS with continuous phase shift, even based on 1-bit discrete phase shift. Furthermore, \cite{DBLP:journals/corr/abs-1810-06934} and \cite{DBLP:conf/globecom/HuangAZDY18} discussed the joint AP power allocation and IRS phase-shift optimization to maximize system energy and spectrum efficiency, where the user has a minimum transmission rate constraint and the AP has a total transmit power constraint. Since the presented problem is non-convex, the gradient descent (GD) based AP power allocation algorithm and fractional programming (FP) based IRS phase shift algorithm were proposed therein. For the IRS assisted wireless communication system, Han et al. \cite{DBLP:journals/tvt/HanTJWM19} analyzed and obtained a compact approximation of system ergodic capacity and then, based on statistical channel information and approximate traversal capacity, the optimal IRS phase shift was proved. The authors also derived the required quantized bits of the IRS discrete phase shift system to obtain an acceptable ergodic capacity degradation. In \cite{DBLP:journals/corr/abs-1904-10136}, a new IRS hardware architecture was presented and then, based on compressed sensing and deep learning, two reflection beamforming methods were proposed with different algorithm complexity and channel estimation training overhead. Similar to \cite{DBLP:journals/corr/abs-1904-10136}, Huang \cite{DBLP:conf/spawc/HuangAYD19} proposed a deep learning based algorithm to maximize the received signal strength for IRS-assisted indoor wireless communication environment. Some recent studies about the IRS assisted wireless communications could be found in \cite{DBLP:conf/icc/TanSJP16, DBLP:conf/infocom/TanSKJ18, DBLP:journals/corr/abs-1907-03133, DBLP:journals/corr/abs-1907-06002, DBLP:journals/icl/DingP20, DBLP:journals/tcom/ZhuHWND21}, and they mostly focused on the IRS assisted millimeter band or non-orthogonal multiple access (NOMA) based wireless communications.

On the other hand, for the IRS assisted CR networks, considering that all nodes were equipped with a single antenna, in \cite{DBLP:journals/icl/GuanWZ20}, an IRS was deployed to assist in the spectrum sharing between a primary user (PU) link and a secondary user (SU) link. For which, authors aimed to maximize the achievable SU rate subject to a given SINR target for the PU link, by jointly optimizing the SU transmit power and IRS reflect beamforming. Different from \cite{DBLP:journals/icl/GuanWZ20}, the CR system consists of multiple SUs and a secondary access point (SAP) with multiple antennas, as introduced in \cite{2020Robust}. Based on both the bounded channel state information (CSI) error model and statistical CSI error model for PU-related channels, robust beamforming design was investigated. Specifically, the transmit precoding matrix at the SAP and phase shifts at the IRS were jointly optimized to minimize the total transmit power at the SAP, subject to the quality of service of SUs, the limited interference imposed on the PU and unit-modulus of the reflective beamforming. In addition, in \cite{DBLP:conf/icc/YuanLJ0L20} and \cite{DBLP:journals/tcom/YuanLJFL21}, Yuan et al. considered a CR system including multiple PUs and single SU, and maximized the transmission rate of SUs under the constraints of the maximum transmitting power at the SAP and the interference leakage at the PUs. Moreover, the research result was extended to the case with imperfect CSI. It is noted that the system mentioned in \cite{DBLP:conf/icc/YuanLJ0L20} contained only one IRS, while multiple IRSs were introduced in \cite{DBLP:journals/tcom/YuanLJFL21}. Furthermore, in \cite{DBLP:conf/spawc/XuYS20}, the beamforming vectors at the base station and the phase shift matrix at the IRS are jointly optimized for maximization of the sum rate of the secondary system based on the network with multiple PUs and SUs, and the suboptimal solution was obtained by the AO-based algorithm. Considering the same scenario, \cite{2020Reconfigurable} minimized the transmit power at the SAP equipped with multiple antennas and the AO-based algorithm was proposed as well. Furthermore, in \cite{DBLP:journals/tcom/XuYSNS20}, an IRS-assisted secondary network employed an full-duplex base station for serving multiple half-duplex downlink and uplink users simultaneously. The downlink transmit beamforming vectors as well as the uplink receive beamforming vectors at the full-duplex base station, the transmit power of the uplink users, and the phase shift matrix at the IRS are jointly optimized for maximization of the total sum rate of the secondary system. The design task is formulated as a non-convex optimization problem taking into account of the imperfect CSI of the PUs and their maximum interference tolerance, for which, an iterative block coordinate descent (BCD)-based algorithm was developed. Combining IRS with physical layer security, \cite{DBLP:journals/sensors/XiaoDW20 } aimed to solve the security issue of CR networks. Specifically, an IRS-assisted MISO CR wiretap channel was studied. To maximize the secrecy rate of SUs subject to a total power constraint for the transmitter and interference power constraint for a single antenna PU, an AO algorithm is proposed to jointly optimize the transmit covariance at transmitter and phase shift coefficient at IRS by fixing the other as constant.

Note that, all the existing research about the IRS-assisted CR network only consider that PUs and SUs are equipped with a single antenna \cite{DBLP:journals/icl/GuanWZ20, 2020Robust, DBLP:conf/icc/YuanLJ0L20, DBLP:journals/tcom/YuanLJFL21, DBLP:conf/spawc/XuYS20, 2020Reconfigurable, DBLP:journals/tcom/XuYSNS20, DBLP:journals/sensors/XiaoDW20}. However, in order to further improve the performance of the wireless systems, multi-antenna enabled technologies are adopted by many commercial standards, such as the IEEE 802.11ax, LTE and the fifth generation (5G) mobile networks. Therefore, it is imperative to study the IRS assisted CR system with multiple-transmit and multiple-receive antennas, and this is the focus of this work. To the best of our knowledge, there is only one work in \cite{DBLP:journals/tvt/ZhangWTJSP20} that is similar to ours but with single PU. In specific, the author proposed an IRS assisted CR system which includes an SAP, a PU and multiple SUs. It should be noted that the proposed algorithm in \cite{DBLP:journals/tvt/ZhangWTJSP20} could not be extended to the general scenario with multiple PUs, namely, our work in this paper is more universal and the scenario studied in \cite{DBLP:journals/tvt/ZhangWTJSP20} can be treated only as a special case of our work. Specifically, the contributions of this paper are summarized as follows.

\begin{itemize}
  \item In this paper, an IRS-assisted downlink multiple-input multiple-output (MIMO) CR system is proposed, that is, an SAP communicates with multiple SUs with the assistance of an IRS but without affecting multiple PUs in the primary network. We aim at maximizing the achievable WSR of SUs by jointly optimizing the transmit precoding matrix at the SAP and the reflecting coefficients at the IRS, subject to a total transmit power constraint at the SAP and interference temperature constraints at PUs. To deal with the complex objective function, the problem is reformulated by employing the well-known weighted minimum mean-square error (WMMSE) method and an AO-based algorithm is proposed. That is, for the auxiliary matrix, decoding matrices, SAP precoding matrix and IRS reflection coefficient matrix, one of them is iteratively obtained while keeping the others fixed, and the process continues until convergence.

  \item In addition, for the scenario with a single PU, i.e., the same scenario as \cite{DBLP:journals/tvt/ZhangWTJSP20}, a lower complexity algorithm is proposed. Since the algorithm proposed in \cite{DBLP:journals/tvt/ZhangWTJSP20} involves the inverse operation of the matrix in the iterative process for SAP precoding matrix optimization under given auxiliary matrix, decoding matrix and IRS reflection coefficients, whose complexity is very high. In this paper, the special structure of the matrices is leveraged to remove the inverse operation. Further, without impacting the performance, the algorithm with lower complexity is presented by the Lagrangian dual decomposition and successive convex approximation (SCA) method, and the optimal solution of the subproblem is obtained.

  \item Finally, some numerical simulations have been provided to demonstrate that the proposed algorithm outperforms other benchmark schemes. Note that, simulation results include two parts which are corresponding to the general scenario with multiple PUs and the special scenario with only one PU.
\end{itemize}

The rest of this paper is organized as follows. In Section \ref{section_2}, the system model and the considered optimization problem are presented. In Section \ref{section_3}, the considered problem is discussed and solved, and an AO-based algorithm is proposed. Then the discussion is extended to a special scenario with only one PU in Section \ref{section_4} and an AO-based algorithm with lower complexity is proposed therein. The simulation results are presented in Section \ref{section_5} and we conclude at last.

\emph{Notation:} We use uppercase boldface letters for matrices and lowercase boldface letters for vectors. $\mathbb{E}\{\bullet\}$ stands for the statistical expectation for random variables, and $|\bullet|$, $arg(\bullet)$, $\Re \{ \bullet \}$ and $(\bullet)^*$ denote the absolute value, the argument, the real part and the conjugate of a complex number, respectively. $\det(\bullet)$ and $Tr(\bullet)$ indicate the determinant and trace of a matrix, respectively, whereas $(\bullet)^T$, $(\bullet)^H$, $(\bullet)^{-1}$ and $(\bullet)^\dag$ represent the transpose, conjugate transpose, inverse and pseudo-inverse of a matrix, respectively. In addition, $\mathbf{I}$ denotes the identity matrix with appropriate size, and $diag(\bullet)$ represents a diagonal matrix whose diagonal elements are from a vector. $\mathbf{A} \succcurlyeq 0$ and $\mathbf{A} \succ 0$ indicate that $\mathbf{A}$ is positive semi-definite and positive definite matrix.

\section{System Model and The Problem}\label{section_2}
In this section, firstly, we present the system model of the intelligent reflecting surface (IRS)-assisted downlink multiple-input multiple-output (MIMO) cognitive radio (CR) system, termed as the IRS-MIMO-CR system. Then, we illustrate the signal model for our considered system, which includes the channel model and IRS reflecting model. Note that, as per in \cite{DBLP:conf/globecom/WuZ18} and \cite{DBLP:journals/twc/WuZ19}, the signals that are reflected by the IRS multi-times are ignored due to significant path loss. Moreover, to characterize the performance limit of the considered IRS-assisted secure communication system, the quasi-static flat-fading channel model is adopted herein and all the CSI are perfectly known at the SAP \cite{DBLP:journals/jsac/ZhangZ20a}. Finally, we formulate the discussed optimization problem.

\subsection{System Model}
\begin{figure}[tbp]
    \centering
    \includegraphics[scale=0.5]{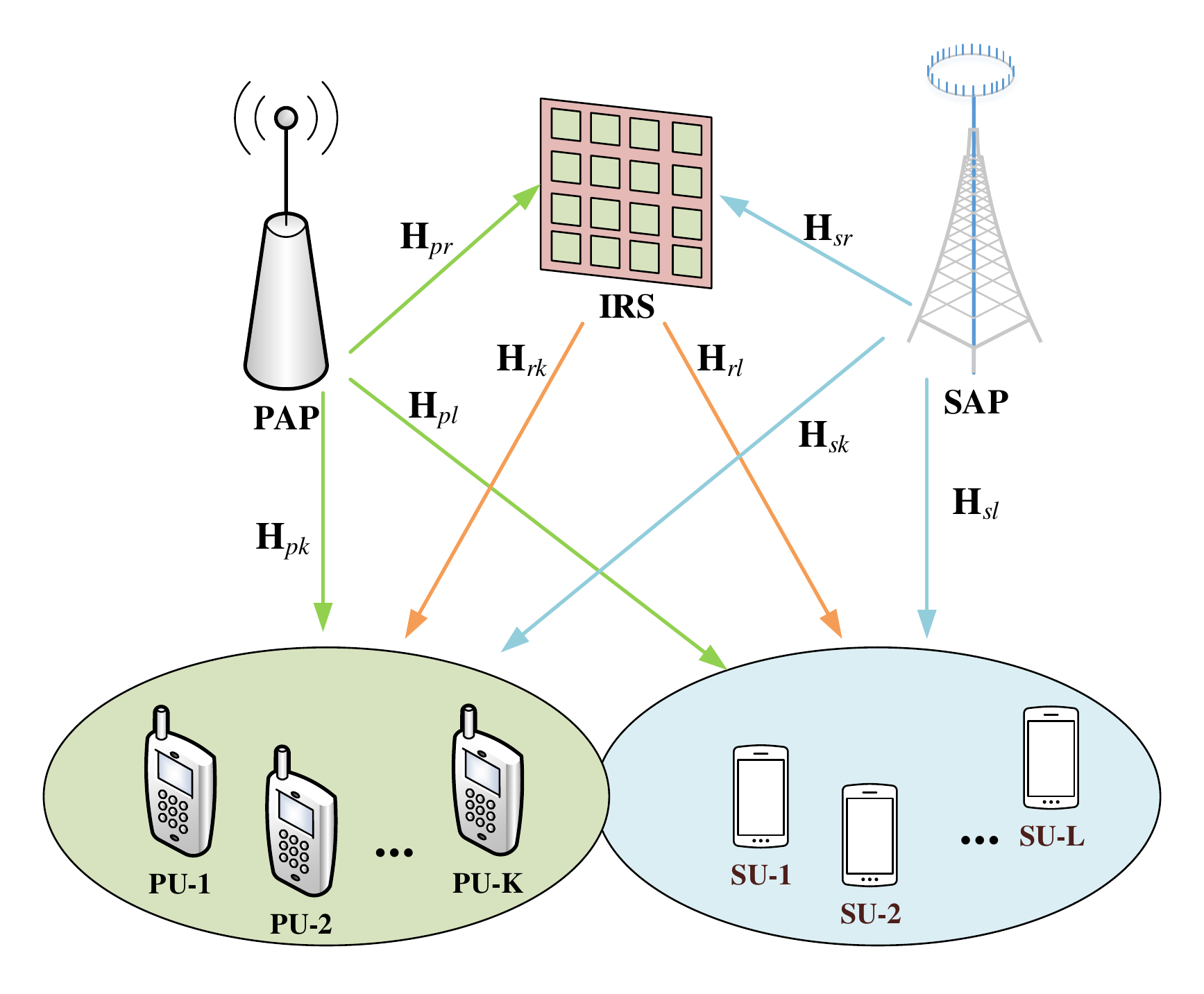}
    \caption{\quad System Model.}
    \label{fig1}
\end{figure}

We consider the IRS-MIMO-CR system, as shown in Fig.~\ref{fig1}, where a secondary access point (SAP) serves multiple secondary users (SUs) without affecting the communications between primary access point (PAP) and multiple primary users (PUs) in the primary network. All nodes are equipped with multiple antennas and the number of antennas at the PAP, PUs, SAP and SUs are $N_{PA}$, $N_{PU}$, $N_{SA}$ and $N_{SU}$, respectively. Denote the sets of PUs and SUs as ${\cal K} = \left\{ {1,2,...,K} \right\}$ and ${\cal L} = \left\{ {1,2,...,L} \right\}$, respectively. In addition, an IRS composed of $M$ passive elements, which are denoted as the set ${\cal M} = \{ 1,2,...,M\}$, is installed on a surrounding wall to assist the communications between the SAP and SUs. The IRS has a smart controller, who has the capability of dynamically adjusting the phase shift of each reflecting element based on the propagation environment learned through periodic sensing \cite{DBLP:conf/globecom/WuZ18}.

The number of data streams for each SU is assumed as $d$, satisfying $1 \le d \le \min \left\{N_{SA}, N_{SU}\right\}$. The transmit signal from the SAP is given by
\begin{equation}\label{eq_sig_tx}
\boldsymbol{x} = \sum\limits_{l = 1}^L {\mathbf{F}_l}{\boldsymbol{s}_l} \; ,
\end{equation}
where $\boldsymbol{s}_l \in {\mathbb{C}^{d \times 1}}, \forall l \in {\cal L}$ is the $d \times 1$ data symbol vector designated for the $l$th SU satisfying $\mathbb{E}\left[ \boldsymbol{s}_l \boldsymbol{s}_l^H \right] = \mathbf{I}$ and $\mathbb{E}\left[ \boldsymbol{s}_i \boldsymbol{s}_j^H \right] = \boldsymbol{0},\forall i,j \in {\cal L},i \ne j$. In addition, $\mathbf{F}_l \in \mathbb{C}^{N_{SA} \times d}, \forall l \in {\cal L}$ is the linear precoding matrix used by the SAP for the $l$th SU. As a result, the received signals, which are transmitted from SAP, at the $k$th PU and $l$th SU are given by
\begin{equation}\label{eq_sig_rx_k}
\begin{split}
\boldsymbol{y}_k &= \boldsymbol{r}_k + \mathbf{H}_{sk}\boldsymbol{x} + \mathbf{H}_{rk}\mathbf{\Theta }\mathbf{H}_{sr}\boldsymbol{x} + \boldsymbol{n}_k \\
&= \boldsymbol{r}_k + \left( \mathbf{H}_{sk} + \mathbf{H}_{rk}\mathbf{\Theta }\mathbf{H}_{sr} \right)\boldsymbol{x} + \boldsymbol{n}_k \\
&= \boldsymbol{r}_k + \mathbf{G}_{sk}\left( \mathbf{\Theta } \right)\boldsymbol{x} + \boldsymbol{n}_k,\\
\end{split}\;
\end{equation}

\begin{equation}\label{eq_sig_rx_l}
\begin{split}
\boldsymbol{y}_l &= \mathbf{H}_{sl}\boldsymbol{x} + \mathbf{H}_{rl}\mathbf{\Theta }\mathbf{H}_{sr}\boldsymbol{x} + \boldsymbol{n}_l \\
&= \left(\mathbf{H}_{sl} + \mathbf{H}_{rl}\mathbf{\Theta }\mathbf{H}_{sr}\right)\boldsymbol{x} + \boldsymbol{n}_l \\
&= \mathbf{G}_{sl}\left( \mathbf{\Theta } \right)\boldsymbol{x} + \boldsymbol{n}_l, \\
\end{split} \;
\end{equation}
where $\boldsymbol{r}_k \in \mathbb{C}^{N_{PU} \times 1}, \forall k \in {\cal K}$ stands for the signal which is transmitted from PAP and received by the $k$th PU. Moreover, the baseband equivalent channels from the SAP to the $k$th PU, from the SAP to the $l$th SU, from the SAP to the IRS, from the IRS to the $k$th PU, from the IRS to the $l$th SU are modelled by matrices $\mathbf{H}_{sk} \in \mathbb{C}^{{N_{PU}} \times {N_{SA}}}$, $\mathbf{H}_{sl} \in \mathbb{C}^{{N_{SU}} \times {N_{SA}}}$, $\mathbf{H}_{sr} \in \mathbb{C}^{M \times {N_{SA}}}$, $\mathbf{H}_{rk} \in \mathbb{C}^{{N_{PU}} \times M}$ and $\mathbf{H}_{rl} \in \mathbb{C}^{{N_{SU}} \times M}$, respectively. Let $\mathbf{\Theta} = diag( {\boldsymbol{\theta}} ) \in {\mathbb{C}^{M \times M}}$ denote the diagonal reflection matrix of the IRS, with $\boldsymbol{\theta} = {\left[ \theta _1, \theta _2,...,\theta _M \right]^T} \in {\mathbb{C}^{M \times 1}}$ and $\left| \theta _m \right| = 1, \; \forall m \in {\cal M}$. Therefore, the effective MIMO channel matrix from the SAP to the $k$th PU and $l$th SU receiver is given by ${\mathbf{G}_{sk}}\left( \mathbf{\Theta} \right) = {\mathbf{H}_{sk}} + {\mathbf{H}_{rk}} \mathbf{\Theta} {\mathbf{H}_{sr}}$ and ${\mathbf{G}_{sl}}\left( \mathbf{\Theta} \right) = {\mathbf{H}_{sl}} + {\mathbf{H}_{rl}} \mathbf{\Theta} {\mathbf{H}_{sr}}$, respectively. $\boldsymbol{n}_k \sim CN\left( \boldsymbol{0},\sigma _k^2\mathbf{I} \right)$ represents the additive white Gaussian noise at the $k$th PU, and $\boldsymbol{n}_l \sim CN\left( \boldsymbol{0},\sigma _l^2\mathbf{I} \right)$ is the equivalent noise at the $l$th SU, which captures the joint effect of the received interference from the primary network and thermal noise. $\sigma _k^2$ and $\sigma _l^2$ denote the corresponding average noise power at the $k$th PU and the $l$th SU. By substituting (\ref{eq_sig_tx}) into (\ref{eq_sig_rx_l}), the received signals of the $l$th SU are reformulated as follows
\begin{equation}\label{eq_sig_rx_l_rw}
\boldsymbol{y}_l = \mathbf{G}_{sl}\left( \mathbf{\Theta } \right)\mathbf{F}_l\boldsymbol{s}_l + \sum\limits_{i = 1,i \ne l}^L \mathbf{G}_{sl}\left( \mathbf{\Theta } \right)\mathbf{F}_i\boldsymbol{s}_i  + \boldsymbol{n}_l \; .
\end{equation}
Denotes the collection of precoding matrixs used by the SAP as $\mathbf{F} = \left\{ \mathbf{F}_1, \mathbf{F}_2,...,\mathbf{F}_L \right\}$. Hence, the transmit data rate (bit/s/Hz) of the $l$th SU is written as follows
\begin{equation}\label{eq_r_l}
{R_l}\left( {\mathbf{F},\mathbf{\Theta } } \right) = {\log_2}\det \left( {\mathbf{I} + \mathbf{G}_{sl}\left( \mathbf{\Theta }  \right)\mathbf{F}_l\mathbf{F}_l^H\mathbf{G}_{sl}^H\left( \mathbf{\Theta } \right)\mathbf{J}_l^{ - 1}} \right) \;,
\end{equation}
where $\mathbf{J}_l = \sum\limits_{i = 1,i \ne l}^L {\mathbf{G}_{sl} \left( \mathbf{\Theta} \right) \mathbf{F}_i \mathbf{F}_i^H \mathbf{G}_{sl}^H \left( \mathbf{\Theta } \right)}  + \sigma _l^2\mathbf{I}$ represents the interference-plus-noise covariance
matrix of the $l$th SU. Moreover, the interference signal power
imposed on the $k$th PU is denoted by
\begin{equation}\label{eq_it_k}
IT_k = \sum\limits_{l = 1}^L {Tr\left(  {\mathbf{G}_{sk}\left( \mathbf{\Theta } \right)\mathbf{F}_l\mathbf{F}_l^H\mathbf{G}_{sk}^H\left( \mathbf{\Theta} \right)} \right)} \; .
\end{equation}

\subsection{Problem Formulation}
As mentioned earlier, we discuss the joint optimization of the transmit precoding matrix at the SAP and the reflection coefficients at the IRS to maximize the achievable WSR of SUs, subject to the total transmit power constraint at the SAP, interference constraints at PUs and the reflection coefficient constraint at the IRS. Thus we have the following OP1,
\begin{equation}\label{eq_OP1}
\begin{split}
 \mathop {\max }\limits_{\mathbf{F},\mathbf{\Theta}}\;&\sum\limits_{l = 1}^L {{\omega _l}{R_l}\left( \mathbf{F},\mathbf{\Theta} \right)}  \\
s.t.\;&C1:\sum\limits_{l = 1}^L { Tr \left( \mathbf{F}_l^H\mathbf{F}_l \right) }\le P_{\max} \\
      &C2:\sum\limits_{l = 1}^L { Tr \left(\mathbf{G}_{sk}\left( \mathbf{\Theta } \right)\mathbf{F}_l\mathbf{F}_l^H\mathbf{G}_{sk}^H\left( \mathbf{\Theta} \right) \right) } \le {\Gamma_k},\forall k \in {\cal K} \\
      &C3:\left| \mathbf{\Theta}_{mm} \right| = 1,\forall m \in {\cal M} \\
\end{split} \; .
\end{equation}
Herein, C1 characterizes the total transmit power constraint at the SAP, C2 defines the interference constraints at PUs, and C3 represents the IRS reflecting coefficient constraint, $\Gamma_k$ denotes the maximum received interference power at the PU $k$. It is evident that OP1 is a non-convex nonlinear programming with coupled variables $\mathbf{F}$ and $\mathbf{\Theta}$ and also the uni-modular constraint on each reflection coefficient $\mathbf{\Theta}_{mm}$, which makes it difficult to solve. Therefore, in the sequel, we pursue the suboptimal approach to handle OP1.

\section{Alternating Optimization based Joint Optimization Algorithm}\label{section_3}
In this section, a suboptimal algorithm is proposed to solve OP1. As aforementioned, the formulated OP1 is a non-convex nonlinear programming. Therefore, we first transform OP1 into a more tractable one, which allows the decoupling of precoding matrices and the reflection coefficient matrix. Then, alternating optimization (AO) algorithm is proposed to solve the transformed problem.

\subsection{Reformulation of the Original Problem}
To deal with the complex objective function, we reformulate OP1 by employing the well-known WMMSE \cite{DBLP:journals/tsp/ShiRLH11} method. Specifically, the linear decoding matrix is applied to estimate the signal vector for each SU, which is denoted by
\begin{equation}\label{eq_sig_l}
\boldsymbol{\hat s}_l = \mathbf{U}_l^H\boldsymbol{y}_l,\forall l \in {\cal L}
\end{equation}
where $\mathbf{U}_l \in \mathbb{C}^{{N_{SU}} \times d}$ is the decoding matrix for the $l$th SU. Then, the MSE matrix for the $l$th SU is given by
\begin{equation}\label{eq_MSE}
\begin{split}
\mathbf{E}_l &= \mathbb{E}_{\boldsymbol{s},\boldsymbol{n}} \left[ \left( \boldsymbol{\hat s}_l - \boldsymbol{s}_l \right) \left( \boldsymbol{\hat s}_l - \boldsymbol{s}_l \right)^H \right] \\
&= \left( {\mathbf{U}_l^H \mathbf{G}_{sl} \left( \mathbf{\Theta } \right) \mathbf{F}_l - \mathbf{I}} \right){\left( {\mathbf{U}_l^H{\mathbf{G}_{sl}}\left( \mathbf{\Theta } \right){\mathbf{F}_l} - \mathbf{I}} \right)^H} \\
&+ \sum\limits_{i = 1,i \ne l}^L {\mathbf{U}_l^H\mathbf{G}_{sl}\left( \mathbf{\Theta} \right) {\mathbf{F}_i} \mathbf{F}_i^H \mathbf{G}_{sl}^H \left( \mathbf{\Theta } \right) \mathbf{U}_l} \\
&+ \sigma_l^2\mathbf{U}_l^H\mathbf{U}_l,\forall l \in {\cal L} \\
\end{split}
\end{equation}
where $\boldsymbol{s}$ and $\boldsymbol{n}$ denote the collections of data symbols and noise vectors of all SUs, respectively.
Denoting the sets of decoding matrices as $\mathbf{U} = \left\{ \mathbf{U}_l, \; \forall l \in {\cal L} \right\}$ and introducing a set of auxiliary matrices $\mathbf{W} = \left\{ \mathbf{W}_l \succcurlyeq \mathbf{0}, \; \forall l \in {\cal L} \right\}$, OP1 can be reformulated as the following OP2
\begin{equation}\label{eq_OP2}
\begin{split}
\mathop {\max }\limits_{\mathbf{W},\mathbf{U},\mathbf{F},\mathbf{\Theta }} \;& \sum\limits_{l = 1}^L {\omega_l h_l \left( \mathbf{W},\mathbf{U},\mathbf{F},\mathbf{\Theta } \right)}  \\
s.t.\;&C1, \; C2 \; and \; C3 \\
\end{split} \; .
\end{equation}
Herein, ${h_l}\left( \mathbf{W},\mathbf{U},\mathbf{F},\mathbf{\Theta} \right)$ is given by
\begin{equation}\label{eq_h_l}
{h_l}\left( \mathbf{W},\mathbf{U},\mathbf{F},\mathbf{\Theta} \right) = \log_2 \det \left( \mathbf{W}_l \right) - Tr \left( \mathbf{W}_l \mathbf{E}_l \right) + d \; .
\end{equation}
Note that, by iteratively obtaining one set of variables while keeping the others fixed, the objective function in OP2 is much easier to handle with AO. Since the decoding matrices $\mathbf{U}$ and auxiliary matrices $\mathbf{W}$ only appear in ${h_l}\left( \mathbf{W}, \mathbf{U}, \mathbf{F}, \mathbf{\Theta} \right)$, the optimal solution of $\mathbf{U}$ and $\mathbf{W}$ can be obtained by setting the first-order derivative of ${h_l}\left( \mathbf{W}, \mathbf{U}, \mathbf{F}, \mathbf{\Theta} \right)$ with respect to $\mathbf{U}_l$ and $\mathbf{W}_l$ to zero while keeping the other matrices fixed. Hence, the optimal solutions are denoted by
\begin{equation}\label{eq_U_l}
{\mathbf{\hat U}_l} = {\left( \mathbf{J}_l + \mathbf{G}_{sl}\mathbf{F}_l\mathbf{F}_l^H\mathbf{G}_{sl}^H \right)^{-1}} \mathbf{G}_{sl} \mathbf{F}_l \;,
\end{equation}
\begin{equation}\label{eq_W_l}
{\mathbf{\hat W}_l} = \left( \mathbf{\hat E}_l \right)^{-1} \;.
\end{equation}
Herein, given the reflection coefficient matrix $\mathbf{\Theta}$ at the
IRS, $\mathbf{G}_{sl}\left( \mathbf{\Theta} \right)$ is simplistically denoted as $\mathbf{G}_{sl}$. Moreover, $\mathbf{\hat E}_l$ is obtained by inserting $\mathbf{\hat U}_l$ into the MSE matrix of the $l$th SU, yielding
\begin{equation}\label{eq_E_l}
\mathbf{\hat E}_l = \mathbf{I} - \mathbf{F}_l^H \mathbf{G}_{sl}^H {\left( {\mathbf{J}_l + \mathbf{G}_{sl} \mathbf{F}_l \mathbf{F}_l^H \mathbf{G}_{sl}^H } \right)^{-1}} \mathbf{G}_{sl} \mathbf{F}_l
\end{equation}
In the following, the precoding matrices $\mathbf{F}$ and reflection coefficient matrix $\mathbf{\Theta}$ are optimized with given $\mathbf{U}$ and $\mathbf{W}$.

\subsection{Optimization of the precoding matrices}
In this subsection, we discuss the precoding matrices optimization at the SAP given the reflecting coefficients at the IRS, decoding matrices and auxiliary matrices. Hence, by substituting $\mathbf{E}$ into the objective function of OP2 and discarding the constant terms, the precoding matrices optimization problem can be transformed as the following OP3,
\begin{equation}\label{eq_OP3}
\begin{split}
\mathop {\min }\limits_\mathbf{F} \;&\sum\limits_{l = 1}^L Tr \left( \mathbf{F}_l^H \mathbf{X}_0 \mathbf{F}_l \right) - \sum\limits_{l = 1}^L 2Re\left\{ Tr \left( \mathbf{Y}_l^H \mathbf{F}_l^H \right)\right\}  \\
s.t. \;& \sum\limits_{l = 1}^L { Tr \left( {\mathbf{F}_l^H {\mathbf{F}_l} } \right) } \le {P_{\max }} \\
&\sum\limits_{l = 1}^L { Tr \left( {\mathbf{F}_l^H{\mathbf{X}_k}{\mathbf{F}_l} } \right) } \le {\Gamma _k},\forall k \in {\cal K} \\
\end{split} \;.
\end{equation}
In which, $\mathbf{G}_{sk}$ is the abbreviated form of $\mathbf{G}_{sk}\left( \mathbf{\Theta} \right)$ with given the reflection coefficient matrix $\mathbf{\Theta}$ at the IRS. In addition, ${\mathbf{X}_0} = \sum\limits_{m = 1}^L {\omega _m} \mathbf{G}_{sm}^H {\mathbf{U}_m} {\mathbf{W}_m} \mathbf{U}_m^H{\mathbf{G}_{sm}} \succcurlyeq \mathbf{0}$, ${\mathbf{X}_k} = \mathbf{G}_{sk}^H {\mathbf{G}_{sk}} \succcurlyeq \mathbf{0}, \; \forall k \in {\cal K}$ and ${\mathbf{Y}_l} = {\omega _l} {\mathbf{W}_l} \mathbf{U}_l^H {\mathbf{G}_{sl}}, \; \forall l \in {\cal L}$. It is obvious that OP3 is quadratically constrained quadratic programming (QCQP) problem \cite{DBLP:books/cu/BV2014}. Since $\mathbf{X}_0 \succcurlyeq \mathbf{0}$ and $\mathbf{X}_k \succcurlyeq \mathbf{0}$, OP3 is convex and can be easily solved with standard interior-point methods.

\subsection{Optimize the IRS reflecting coefficients}
In this subsection, given the precoding matrices at the SAP, decoding matrices and auxiliary matrices, the optimization of the reflecting coefficient matrix $\mathbf{\Theta}$ at IRS is discussed. Particularly, we have the following OP4:
\begin{equation}\label{eq_OP4}
\begin{split}
\mathop{\min}\limits_\mathbf{\Theta}\;&\sum\limits_{l = 1}^L Tr \left({{\omega_l} {\mathbf{W}_l} \mathbf{U}_l^H {\mathbf{G}_{sl}} \left( \mathbf{\Theta } \right) {\mathbf{Q}_s} \mathbf{G}_{sl}^H \left( \mathbf{\Theta} \right) {\mathbf{U}_l}} \right) \\
&- \sum\limits_{l = 1}^L 2Re\left\{{Tr \left( {{\omega_l} {\mathbf{W}_l} \mathbf{U}_l^H {\mathbf{G}_{sl}} \left( \mathbf{\Theta} \right){\mathbf{F}_l}} \right)}\right\}  \\
s.t. \;& Tr \left( {{\mathbf{G}_{sk}}\left( \mathbf{\Theta } \right){\mathbf{Q}_s} \mathbf{G}_{sk}^H\left( \mathbf{\Theta } \right)} \right) \le {\Gamma _k},\forall k \in {\cal K} \\
\;& \left| \mathbf{\Theta}_{mm} \right| = 1, \; \forall m \in {\cal M} \\
\end{split}
\end{equation}
where ${\mathbf{Q}_s} = \sum\limits_{m = 1}^L {{\mathbf{F}_m} \mathbf{F}_m^H} $. By applying ${\mathbf{G}_{sl}} \left( \mathbf{\Theta } \right) = {\mathbf{H}_{sl}} + {\mathbf{H}_{rl}} \mathbf{\Theta } {\mathbf{H}_{sr}}$, we have
\begin{equation}\label{eq_IRS_reform1}
\begin{split}
{\omega_l} {\mathbf{W}_l} \mathbf{U}_l^H {\mathbf{G}_{sl}} \left( \mathbf{\Theta} \right) {\mathbf{Q}_s} \mathbf{G}_{sl}^H \left( \mathbf{\Theta } \right) {\mathbf{U}_l}=
\;& {\omega_l} {\mathbf{W}_l} \mathbf{U}_l^H {\mathbf{H}_{rl}} \mathbf{\Theta} {\mathbf{H}_{sr}} {\mathbf{Q}_s} \mathbf{H}_{sr}^H {\mathbf{\Theta }^H} \mathbf{H}_{rl}^H {\mathbf{U}_l} \\
+ \;& {\omega _l} {\mathbf{W}_l} \mathbf{U}_l^H {\mathbf{H}_{rl}} \mathbf{\Theta} {\mathbf{H}_{sr}} {\mathbf{Q}_s} \mathbf{H}_{sl}^H {\mathbf{U}_l} \\
+ \;& {\omega _l}{\mathbf{W}_l}\mathbf{U}_l^H{\mathbf{H}_{sl}}{\mathbf{Q}_s}\mathbf{H}_{sr}^H{\mathbf{\Theta }^H}\mathbf{H}_{rl}^H{\mathbf{U}_l}\\
+ \;& {\omega_l} {\mathbf{W}_l} \mathbf{U}_l^H {\mathbf{H}_{sl}} {\mathbf{Q}_s} \mathbf{H}_{sl}^H {\mathbf{U}_l} \\
\end{split}
\end{equation}

\begin{equation}\label{eq_IRS_reform2}
{\omega_l} {\mathbf{W}_l} \mathbf{U}_l^H {\mathbf{G}_{sl}} \left( \mathbf{\Theta} \right) {\mathbf{F}_l}= {\omega_l} {\mathbf{W}_l} \mathbf{U}_l^H {\mathbf{H}_{rl}} \mathbf{\Theta} {\mathbf{H}_{sr}} {\mathbf{F}_l} + {\omega_l} {\mathbf{W}_l} \mathbf{U}_l^H {\mathbf{H}_{sl}} {\mathbf{F}_l}
\end{equation}

\begin{equation}\label{eq_IRS_reform3}
\begin{split}
{\mathbf{G}_{sk}} \left( \mathbf{\Theta} \right) {\mathbf{Q}_s} \mathbf{G}_{sk}^H \left( \mathbf{\Theta} \right) = & {\mathbf{H}_{rk}} \mathbf{\Theta} {\mathbf{H}_{sr}} {\mathbf{Q}_s} \mathbf{H}_{sr}^H {\mathbf{\Theta}^H} \mathbf{H}_{rk}^H + {\mathbf{H}_{rk}} \mathbf{\Theta} {\mathbf{H}_{sr}} {\mathbf{Q}_s} \mathbf{H}_{sk}^H \\
+ & {\mathbf{H}_{sk}} {\mathbf{Q}_s} \mathbf{H}_{sr}^H {\mathbf{\Theta}^H} \mathbf{H}_{rk}^H + {\mathbf{H}_{sk}} {\mathbf{Q}_s} \mathbf{H}_{sk}^H \\
\end{split}
\end{equation}
Based on the above conversions, by discarding the constant terms, OP4 can be transformed into the following formulation,
\begin{equation}\label{eq_IRS_reform4}
\begin{split}
\mathop{\min}\limits_\mathbf{\Theta} \;& Tr \left( {{\mathbf{B}_0}\mathbf{\Theta C}{\mathbf{\Theta}^H}} \right)+ 2Re \left\{ Tr \left( {\mathbf{D}_0^H{\mathbf{\Theta }^H}} \right)\right\} \\
s.t. \;& Tr \left( {{\mathbf{B}_k}\mathbf{\Theta C}{\mathbf{\Theta }^H}} \right) + 2Re\left\{ Tr \left( {\mathbf{D}_k^H{\mathbf{\Theta }^H}} \right) \right\} \le {{\tilde \Gamma }_k},\forall k \in {\cal K} \\
& \left| {{\mathbf{\Theta}_{mm}}} \right| = 1,\forall m \in {\cal M} \\
\end{split} \;.
\end{equation}
Herein, ${\mathbf{B}_0} = \sum\limits_{l = 1}^L {{\omega _l} \mathbf{H}_{rl}^H{\mathbf{U}_l}{\mathbf{W}_l}\mathbf{U}_l^H{\mathbf{H}_{rl}}} \succcurlyeq \mathbf{0}$, ${\mathbf{B}_k} = \mathbf{H}_{rk}^H {\mathbf{H}_{rk}} \succcurlyeq \mathbf{0}$, $\mathbf{C} = {\mathbf{H}_{sr}}{\mathbf{Q}_s}\mathbf{H}_{sr}^H \succcurlyeq \mathbf{0}$, ${\mathbf{D}_0} = \sum\limits_{l = 1}^L {{\omega _l}{\mathbf{H}_{sr}}{\mathbf{Q}_s}\mathbf{H}_{sl}^H{\mathbf{U}_l}{\mathbf{W}_l}\mathbf{U}_l^H{\mathbf{H}_{rl}}}  - \sum\limits_{l = 1}^L {{\omega _l} {\mathbf{H}_{sr}} {\mathbf{F}_l} {\mathbf{W}_l} \mathbf{U}_l^H{\mathbf{H}_{rl}}}$, ${\mathbf{D}_k} = {\mathbf{H}_{sr}} {\mathbf{Q}_s} \mathbf{H}_{sk}^H {\mathbf{H}_{rk}}$ and ${\tilde \Gamma _k} = {\Gamma _k} - {\rm{Tr}}\left( {{\mathbf{H}_{sk}}{\mathbf{Q}_s}\mathbf{H}_{sk}^H} \right)$. Since $\mathbf{\Theta} = diag( {\boldsymbol{\theta}} )$ is a diagonal matrix, by adopting the matrix identity in \cite{2017Matrix}, it follows that
\begin{equation}\label{eq_IRS_reform5}
\begin{split}
Tr \left( {{\mathbf{B}_0}\mathbf{\Theta C}{\mathbf{\Theta }^H}} \right){\rm{ = }}{{\boldsymbol{\theta }}^H}\left( {{\mathbf{B}_0} \odot \mathbf{C}} \right){\boldsymbol{\theta}} \\
Tr \left( {{\mathbf{B}_k}\mathbf{\Theta C}{\mathbf{\Theta }^H}} \right){\rm{ = }}{{\boldsymbol{\theta }}^H}\left( {{\mathbf{B}_k} \odot \mathbf{C}} \right){\boldsymbol{\theta }} \\
\end{split} \; .
\end{equation}

Further, denote ${{\boldsymbol{d}}_0} = {\left[ {{{\left[ {{\mathbf{D}_0}} \right]}_{1,1}},{{\left[ {{\mathbf{D}_0}} \right]}_{2,2}},...,{{\left[ {{\mathbf{D}_0}} \right]}_{M,M}}} \right]^T}$ and ${{\boldsymbol{d}}_k} = {\left[ {{{\left[ {{\mathbf{D}_k}} \right]}_{1,1}},{{\left[ {{\mathbf{D}_k}} \right]}_{2,2}},...,{{\left[ {{\mathbf{D}_k}} \right]}_{M,M}}} \right]^T}$ as the collections of diagonal elements of $\mathbf{D}_0$ and $\mathbf{D}_k$. We thus have
\begin{equation}\label{eq_IRS_reform6}
\begin{split}
Tr \left( {\mathbf{D}_0^H{\mathbf{\Theta }^H}} \right){\rm{ = }}{{\boldsymbol{\theta }}^H}{\boldsymbol{d}}_0^*  \; , \;
Tr \left( {\mathbf{D}_k^H{\mathbf{\Theta }^H}} \right){\rm{ = }}{{\boldsymbol{\theta }}^H}{\boldsymbol{d}}_k^* \\
\end{split} \;.
\end{equation}
Therefore, the problem (\ref{eq_IRS_reform4}) can be simplified as follows
\begin{equation}\label{eq_IRS_reform7}
\begin{split}
\mathop{\min}\limits_{\boldsymbol{\theta }} \;& {{\boldsymbol{\theta }}^H}{\mathbf{\Upsilon} _0}{\boldsymbol{\theta }} + 2Re \left\{ {{{\boldsymbol{\theta }}^H}{\boldsymbol{d}}_0^*} \right\} \\
s.t. \;& {{\boldsymbol{\theta }}^H}{\mathbf{\Upsilon} _k}{\boldsymbol{\theta }}+ 2Re\left\{ {{{\boldsymbol{\theta }}^H}{\boldsymbol{d}}_k^*} \right\} \le {{\tilde \Gamma }_k}, \; \forall k \in {\cal K} \\
& \left| {\boldsymbol{\theta}_m} \right| = 1,\;\forall m \in {\cal M} \\
\end{split} \;.
\end{equation}
Herein, since $\mathbf{B}_0 \succcurlyeq \mathbf{0}$, $\mathbf{B}_k \succcurlyeq \mathbf{0}$ and $\mathbf{C} \succcurlyeq \mathbf{0}$, we have $\mathbf{\Upsilon}_0 = \mathbf{B}_0 \odot \mathbf{C}^T \succcurlyeq \mathbf{0}$ and $\mathbf{\Upsilon}_k = \mathbf{B}_k \odot \mathbf{C}^T \succcurlyeq \mathbf{0}$. However, due to the non-convexity of the uni-modulus constraint on each reflection coefficient $\boldsymbol{\theta}_m$, the problem (\ref{eq_IRS_reform7}) is non-convex. Hence, penalty function (PF) and successive convex approximation (SCA) are adopted. Specifically, introducing the slack factor $\lambda \le 0$, the problem (\ref{eq_IRS_reform7}) can be reformulated as follows
\begin{equation}\label{eq_IRS_reform8}
\begin{split}
\mathop {\min }\limits_{\boldsymbol{\theta }} \;& {{\boldsymbol{\theta }}^H}{\mathbf{\Upsilon}_0}{\boldsymbol{\theta }} + 2Re \left\{ {{{\boldsymbol{\theta }}^H}{\boldsymbol{d}}_0^*} \right\} - \lambda {{\boldsymbol{\theta }}^H}{\boldsymbol{\theta }} \\
s.t. \;& {{\boldsymbol{\theta }}^H}{\mathbf{\Upsilon}_k}{\boldsymbol{\theta }} + 2Re \left\{ {{{\boldsymbol{\theta }}^H} {\boldsymbol{d}}_k^*} \right\} \le {{\tilde \Gamma }_k}, \; \forall k \in {\cal K} \\
&\left| {{\boldsymbol{\theta}_m}} \right| \le 1, \; \forall m \in {\cal M} \\
\end{split} \;.
\end{equation}
The term $\lambda \boldsymbol{\theta}^H \boldsymbol{\theta}$ could ensure that the uni-modulus constraint $\left| \boldsymbol{\theta}_m \right| = 1, \; \forall m \in {\cal M}$ is hold at the optimal solution when $\lambda \to + \infty$. Note that, the objective function of the problem (\ref{eq_IRS_reform8}) is the sum of a convex function and a concave function which means (\ref{eq_IRS_reform8}) is non-convex. Hence, the SCA-based algorithm \cite{DBLP:journals/tsp/ScutariFL17, DBLP:journals/tsp/ScutariFLSS17} is used to handle (\ref{eq_IRS_reform8}) and the concave part of the objective function is approximated by its first order Taylor expansion. In specific, given the initial point $\boldsymbol{\theta}^{(n)}$ and by discarding the constant terms, the sub-problem can be denoted by
\begin{equation}\label{eq_IRS_reform9}
\begin{split}
\mathop {\min}\limits_{\boldsymbol{\theta }} \;& {{\boldsymbol{\theta }}^H}{\mathbf{\Upsilon}_0}{\boldsymbol{\theta }} + 2Re \left\{ {{{\boldsymbol{\theta }}^H}{\boldsymbol{d}}_0^*} \right\} - 2\lambda Re \left\{ \boldsymbol{\theta}^H \boldsymbol{\theta}^{(n)} \right\} \\
s.t.\;& {{\boldsymbol{\theta}}^H}{\mathbf{\Upsilon}_k}{\boldsymbol{\theta }} + 2Re \left\{ {{{\boldsymbol{\theta }}^H}{\boldsymbol{d}}_k^*} \right\} \le {{\tilde \Gamma }_k}, \; \forall k \in {\cal K} \\
&\left| {{\boldsymbol{\theta}_m}} \right| \le 1, \; \forall m \in {\cal M} \\
\end{split} \;.
\end{equation}
Now, (\ref{eq_IRS_reform9}) is convex and can be solved by standard interior-point methods \cite{DBLP:books/cu/BV2014}. Therefore, (\ref{eq_IRS_reform8}) can be tackled by solving a series of convex problems iteratively and the details are summarized in Algorithm 1 as below.
\begin{table}[H]
    \centering
    \begin{tabular}{l}
        \hline
        \textbf{Algorithm 1:} SCA-based Algorithm to Solve (\ref{eq_IRS_reform8})\\
        \hline
        S1: Initialize $\boldsymbol{\theta}^{(0)}$, $\varepsilon >0$, $n = 0$, calculate the objective value of \\
        \quad (\ref{eq_IRS_reform8}) as $z \left( \boldsymbol{\theta}^{(0)} \right)$;\\
        S2: Given $\boldsymbol{\theta}^{(n)}$, obtain $\boldsymbol{\hat{\theta}}^{(n)}$ by solving problem (\ref{eq_IRS_reform9}) with CVX;\\
        S3: If $\left| z\left( \boldsymbol{\hat \theta}^{(n)} \right) - z\left( \boldsymbol{\theta}^{(n)} \right) \right| > \varepsilon$, set $z\left( \boldsymbol{\theta}^{(n+1)} \right) = z\left( \boldsymbol{\hat \theta}^{(n)} \right)$, \\
        \quad $\boldsymbol{\theta}^{(n+1)}=\boldsymbol{\hat{\theta}}^{(n)}$, $n=n+1$, go back to S2; else set $\boldsymbol{\hat{\theta}}=\boldsymbol{\hat{\theta}}^{(n)}$;\\
        S4: Output $\boldsymbol{\hat{\theta}}$;\\
        \hline
    \end{tabular}
    \label{algor_1}
\end{table}
Based on the above discussion, we formulate the PF-based algorithm to solve the OP4 as the following Algorithm 2.
\begin{table}[H]
    \centering
    \begin{tabular}{l}
        \hline
        \textbf{Algorithm 2:} PF-based Algorithm to Solve OP4\\
        \hline
        S1: Initialize $\boldsymbol{\theta}^{(0)}$, $\varepsilon >0$, $\lambda ^{(0)}$, $n = 0$;\\
        S2: Given $\lambda ^{(n)}$, obtain $\boldsymbol{\hat{\theta}}^{(n)}$ by solving problem (\ref{eq_IRS_reform8}) using Algorithm 1;\\
        S3: If $\sum\limits_{m = 1}^M {{\left( {\left| \boldsymbol{\hat \theta}_m^{(n)} \right| - 1} \right)}^2}  > \varepsilon$, set $\boldsymbol{\theta}^{(n+1)}=\boldsymbol{\hat{\theta}}^{(n)}$, $n=n+1$, \\
        \quad go back to S2; else set $\boldsymbol{\hat{\theta}}=\boldsymbol{\hat{\theta}}^{(n)}$;\\
        S4: Output $\boldsymbol{\hat{\theta}}$;\\
        \hline
    \end{tabular}
    \label{algor_2}
\end{table}
Note that, we adopt the stopping criterion $\sum\limits_{m = 1}^M {{\left( {\left| \boldsymbol{\hat \theta}_m^{(n)} \right| - 1} \right)}^2}  \leq \varepsilon$ to ensure the uni-modulus constraint holds at the optimal solution for Algorithm 2.

\subsection{Overall Algorithm}
In this subsection, the overall algorithm for OP1 is provided. As mentioned, the algorithm is based on alternating optimization, which optimizes the objective function with respect to different subsets of optimization variables in each iteration while the other subsets are fixed. Therefore, it is summarized as the following Algorithm 3.
\begin{table}[H]
    \centering
    \begin{tabular}{l}
        \hline
        \textbf{Algorithm 3:} AO-based Algorithm to Solve OP1\\
        \hline
        S1: Initialize $\mathbf{F} ^{(0)}$, $\boldsymbol{\theta}^{(0)}$, $\varepsilon >0$, $n = 0$, calculate the WSR of all SUs as \\ \quad $R\left( \mathbf{F} ^{(0)}, \mathbf{\Theta} ^{(0)} \right)$;\\
        S2: Given $\mathbf{F} ^{(n)}$ and $\boldsymbol{\theta} ^{(n)}$, obtain $\mathbf{\hat U}^{(n)}$ and $\mathbf{\hat W}^{(n)}$ according to (\ref{eq_U_l}) and \\
        \quad (\ref{eq_W_l});\\
        S3: Given $\mathbf{\hat U}^{(n)}$, $\mathbf{\hat W}^{(n)}$ and $\boldsymbol{\theta}^{(n)}$, obtain $\mathbf{\hat F}^{(n)}$ by solving the OP3 with \\
        \quad CVX;\\
        S4: Given $\mathbf{\hat U}^{(n)}$, $\mathbf{\hat W}^{(n)}$ and $\mathbf{\hat F}^{(n)}$, obtain $\boldsymbol{\hat \theta}^{(n)}$ by solving the OP4 using \\
        \quad Algorithm 2;\\
        S5: If $\left| {R\left( {{\mathbf{\hat F}^{(n)}},{\mathbf{\hat \Theta }^{(n)}}} \right) - R\left( {{\mathbf{F}^{(n)}} , {\mathbf{\Theta }^{(n)}}} \right)} \right| > \varepsilon$, set $\mathbf{F}^{(n+1)} = \mathbf{\hat F}^{(n+1)}$\\
        \quad $\boldsymbol{\theta}^{(n+1)}=\boldsymbol{\hat{\theta}}^{(n)}$, ${R\left( {{\mathbf{F}^{(n+1)}},{\mathbf{\Theta }^{(n+1)}}} \right) = R\left( {{\mathbf{\hat F}^{(n)}} , {\mathbf{\hat \Theta}^{(n)}}} \right)}$, \\
        \quad $n=n+1$, go back to S2; else set $\mathbf{\hat F}=\mathbf{\hat F}^{(n)}$, $\boldsymbol{\hat{\theta}}=\boldsymbol{\hat{\theta}}^{(n)}$;\\
        S6: Output $\mathbf{\hat{F}}$ and $\boldsymbol{\hat{\theta}}$;\\
        \hline
    \end{tabular}
    \label{algor_3}
\end{table}

\noindent where $\mathbf{\hat U}^{(n)}$, $\mathbf{\hat W}^{(n)}$, $\mathbf{\hat F}^{(n)}$ and $\boldsymbol{\hat \theta}^{(n)}$ represent the stable solutions obtained by solving the subproblems in the $n$th iteration, and $R\left( \mathbf{F,\Theta } \right)$ denotes the WSR of all SUs. Since the original problem is bounded and the progress of the alternative optimization is monotonically non-decreasing, thus the above algorithm is surely convergent. Furthermore, we analyze the computational complexity of the proposed Algorithm 3. The complexity of the algorithm mainly depends on Step 3 and Step 4, the complexity of which are $O\left( L {d^3} N_{SA}^3 \right)$ and $O\left( {{T_1}{T_2}{M^3}} \right)$ \cite{DBLP:books/cu/BV2014}, respectively. In which, $T_1$, $T_2$ and $T_3$ denote the iteration numbers of the Algorithm 1, 2 and 3, respectively. Hence, the complexity of the overall algorithm is denoted as $O\left( {{T_3}\left( {L{d^3}N_{SA}^3 + {T_1}{T_2}{M^3}} \right)} \right)$.

\section{The Special Scenario With Only One PU}\label{section_4}
In this section, the IRS-assisted CR network with single PU is discussed and in which, an AO-based algorithm with the lower complexity is proposed. Specifically, let ${\Gamma}_p$ denote the maximum received interference power at the unique PU. $\mathbf{H}_{sp} \in \mathbb{C}^{{N_{PU}} \times {N_{SA}}}$ and $\mathbf{H}_{rp} \in \mathbb{C}^{{N_{PU}} \times M}$ represent the baseband equivalent channels from the SAP and the IRS to the PU, respectively. Therefore, the effective MIMO channel matrix from the SAP to the PU is given by ${\mathbf{G}_{sp}}\left( \mathbf{\Theta} \right) = {\mathbf{H}_{sp}} + {\mathbf{H}_{rp}} \mathbf{\Theta} {\mathbf{H}_{sr}}$. Therefore, OP1 can be simplified as the following OP5,
\begin{equation}\label{eq_OP5}
\begin{split}
\mathop{\max}\limits_{ \mathbf{F},\mathbf{\Theta} } \;& \sum\limits_{l = 1}^L {{\omega_l}{R_l}\left( { \mathbf{F},\mathbf{\Theta} } \right)}  \\
s.t. \;& C1: \sum\limits_{l = 1}^L { Tr\left( { \mathbf{F}_l^H{\mathbf{F}_l} } \right) } \le {P_{\max }} \\
&C2: \sum\limits_{l = 1}^L { Tr \left( {{\mathbf{G}_{sp}}\left( \mathbf{\Theta } \right) {\mathbf{F}_l} \mathbf{F}_l^H \mathbf{G}_{sp}^H\left( \mathbf{\Theta } \right)} \right) } \le {\Gamma _p} \\
&C3:\left| {{\mathbf{\Theta}_{mm}}} \right| = 1, \; \forall m \in {\cal M} \\
\end{split} \;.
\end{equation}

Note that, the model adopted in this section differs from that in the previous section in the number of PUs, which only affects the number of  interference constraints at PUs. Similarly, the OP5 is transformed to the following OP6 with the WMMSE method,
\begin{equation}\label{eq_OP6}
\begin{split}
\mathop{\max} \limits_{\mathbf{W},\mathbf{U},\mathbf{F},\mathbf{\Theta}} \;& \sum\limits_{l = 1}^L {{\omega_l}{h_l}\left( {\mathbf{W},\mathbf{U},\mathbf{F},\mathbf{\Theta}} \right)}  \\
s.t. \;& C1, \; C2 \; and \; C3 \\
\end{split} \;.
\end{equation}
Based on the above, the AO algorithm is adopted again and the optimal solution of $\mathbf{U}$ and $\mathbf{W}$ can be calculated by (\ref{eq_U_l}) and (\ref{eq_W_l}). In the following, the precoding matrices $\mathbf{F}$ and reflection coefficient matrix $\mathbf{\Theta}$ are optimized with given $\mathbf{U}$ and $\mathbf{W}$.

\subsection{Optimization of the precoding matrices}
In this subsection, we discuss the precoding matrices optimization at the SAP for given the reflecting coefficients at the IRS, decoding matrices and auxiliary matrices. Hence, substituting $\mathbf{E}$ into the objective function of OP6 and discarding the constant terms, the precoding matrices optimization problem can be transformed as the following OP7,
\begin{equation}\label{eq_OP7}
\begin{split}
\mathop{\min}\limits_\mathbf{F} \;& \sum\limits_{l = 1}^L {Tr \left( {\mathbf{F}_l^H {\mathbf{X}_0} {\mathbf{F}_l}} \right)}  - \sum\limits_{l = 1}^L { 2Re \left\{ { Tr \left( {\mathbf{Y}_l^H \mathbf{F}_l^H} \right)} \right\}}  \\
s.t. \;& C1:\sum\limits_{l = 1}^L {Tr \left( {\mathbf{F}_l^H{\mathbf{F}_l}} \right) } \le {P_{\max }} \\
&C2:\sum\limits_{l = 1}^L { Tr \left( {\mathbf{F}_l^H{\mathbf{X}_p}{\mathbf{F}_l}} \right) } \le {\Gamma _p} \\
\end{split} \;.
\end{equation}
Herein, $\mathbf{G}_{sp}$ is the abbreviated form of $\mathbf{G}_{sp}\left( \mathbf{\Theta} \right)$ with given the reflection coefficient matrix $\mathbf{\Theta}$ at the IRS. In addition, ${\mathbf{X}_0} = \sum\limits_{m = 1}^L {\omega _m} \mathbf{G}_{sm}^H {\mathbf{U}_m} {\mathbf{W}_m} \mathbf{U}_m^H{\mathbf{G}_{sm}} \succcurlyeq \mathbf{0}$, ${\mathbf{X}_p} = \mathbf{G}_{sp}^H {\mathbf{G}_{sp}} \succcurlyeq \mathbf{0}$ and ${\mathbf{Y}_l} = {\omega _l} {\mathbf{W}_l} \mathbf{U}_l^H {\mathbf{G}_{sl}}, \; \forall l \in {\cal L}$. Same as OP3, since $\mathbf{X}_0 \succcurlyeq \mathbf{0}$ and $\mathbf{X}_p \succcurlyeq \mathbf{0}$, OP7 is QCQP convex optimization problem \cite{DBLP:books/cu/BV2014}, which can be solved by the standard convex solver packages such as CVX. However, the computational complexity is high. In the following, we provide a low-complexity SCA-based algorithm which can obtain the optimal solution of the OP7 by solving a series of simple convex problem with Lagrangian dual decomposition method. For which, we introduce the following proposition.

\textbf{Proposition 1:} Let $f_p\left(\mathbf{F}\right)=\sum\limits_{l = 1}^L {Tr \left( {\mathbf{F}_l^H{\mathbf{X}_p}{\mathbf{F}_l}} \right)}$, ${\mathbf{X}_p} \succcurlyeq \mathbf{0}$, $\mathbf{Z}_p = {\lambda}_p \mathbf{I}$ and ${\lambda}_p$ denotes the maximum eigenvalue of the $\mathbf{X}_p$. Then for $\forall \mathbf{F}$ and given $\mathbf{F}^{(n)}$, there exists
\begin{equation}\label{eq_proposition1}
\begin{split}
{{\tilde f}_p}\left( {\mathbf{F}|{\mathbf{F}^{(n)}}} \right) \;& = \sum\limits_{l = 1}^L {Tr \left( {\mathbf{F}_l^H{\mathbf{Z}_p}{\mathbf{F}_l}} \right)} + \sum\limits_{l = 1}^L {Tr \left( {\mathbf{F}_l^{(n)H}\left( {{\mathbf{Z}_p} - {\mathbf{X}_p}} \right)\mathbf{F}_l^{(n)}} \right)} \\
\;& - \sum\limits_{l = 1}^L { 2Re \left\{ {Tr\left( {\mathbf{F}_l^{(n)H}\left( {{\mathbf{Z}_p} - {\mathbf{X}_p}} \right){\mathbf{F}_l}} \right)} \right\}}\\
\end{split}
\end{equation}
which satisfies the following three conditions:

1) ${\tilde f_p}\left( {{\mathbf{F}^{(n)}}|{\mathbf{F}^{(n)}}} \right) = {f_p}\left( {{\mathbf{F}^{(n)}}} \right)$,

2) ${\left. {{\nabla _{{\mathbf{F}}^*}}{{\tilde f}_p}\left( {\mathbf{F}|{\mathbf{F}^{(n)}}} \right)} \right|_{\mathbf{F} = {\mathbf{F}^{(n)}}}} = {\left. {{\nabla _{{\mathbf{F}}^*}}{f_p}\left( \mathbf{F} \right)} \right|_{\mathbf{F} = {\mathbf{F}^{(n)}}}}$,

3) ${\tilde f_p}\left( {\mathbf{F}|{\mathbf{F}^{(n)}}} \right) \ge {f_p}\left( \mathbf{F} \right)$.

\emph{Proof:} Please see the Appendix \ref{appendix_1}. $\hfill\blacksquare$

Based on Proposition 1, the interference constraint C2 in OP7 can be replaced by the following inequality,
\begin{equation}\label{eq_scene1_inter}
\sum\limits_{l = 1}^L {Tr \left( \mathbf{F}_l^H {\mathbf{Z}_p} {\mathbf{F}_l} \right)} - \sum\limits_{l = 1}^L { 2Re \left\{ Tr\left( {\mathbf{F}_l^{(n)H}\left( {{\mathbf{Z}_p} - {\mathbf{X}_p}} \right){\mathbf{F}_l}} \right) \right\}} \le {\tilde \Gamma_p}.
\end{equation}

Herein, ${\tilde \Gamma _p} = {\Gamma _p} - \sum\limits_{l = 1}^L {Tr\left( {\mathbf{F}_l^{(n)H}\left( {{\mathbf{Z}_p} - {\mathbf{X}_p}} \right) \mathbf{F}_l^{(n)}} \right)}$. Therefore, we have the following problem
\begin{equation}\label{eq_appro_pro}
\begin{split}
\mathop {\min}\limits_\mathbf{F} \;& \sum\limits_{l = 1}^L {Tr \left( {\mathbf{F}_l^H{\mathbf{X}_0}{\mathbf{F}_l}} \right)}  - \sum\limits_{l = 1}^L {2Re \left\{ {Tr \left( {\mathbf{Y}_l^H \mathbf{F}_l^H} \right)} \right\}}  \\
s.t. \;& (28 C1), \; (30) \\
\end{split}
\end{equation}
We note that (\ref{eq_appro_pro}) is convex and thus traditional interior-point methods (IPM) can be used to handle it. However, to avoid high computational complexity of the IPM, herein, a low-complexity algorithm based on Lagrange duality decomposition is presented.

Specifically, assuming the optimal solution of (\ref{eq_appro_pro}) is $\mathbf{\hat F}$, herein, according to whether the power constraint (28 C1) is an active constraint at $\mathbf{\hat F}$, two cases are discussed in the following.

\underline{\emph{Case 1}}: Assuming the power constraint (28 C1) is an inactive constraint at $\mathbf{\hat F}$, problem (\ref{eq_appro_pro}) can be transformed into the following problem,
\begin{equation}\label{eq_scene1_appro_pro}
\begin{split}
\mathop {\min}\limits_\mathbf{F} \;& \sum\limits_{l = 1}^L {Tr \left( {\mathbf{F}_l^H{\mathbf{X}_0}{\mathbf{F}_l}} \right)}  - \sum\limits_{l = 1}^L {2Re \left\{ {Tr \left( {\mathbf{Y}_l^H \mathbf{F}_l^H} \right)} \right\}}  \\
s.t. \;& (30) \\
\end{split}
\end{equation}
Introducing the Lagrange multiplier $\mu$ associated with the interference constraint (\ref{eq_scene1_inter}), the Lagrangian function for problem (\ref{eq_scene1_appro_pro}) can be derived as follows
\begin{equation}\label{eq_scene1_lagrange_func}
\begin{split}
L\left( {\mathbf{F},\mu } \right) \;& = \sum\limits_{l = 1}^L {Tr \left( {\mathbf{F}_l^H \left( {{\mathbf{X}_0} + \mu {\mathbf{Z}_p}} \right){\mathbf{F}_l}} \right)} - \sum\limits_{l = 1}^L { 2Re \left\{ {Tr \left( {\mathbf{Y}_l^H \mathbf{F}_l^H} \right)} \right\}} - \mu {{\tilde \Gamma }_p}  \\
&- \sum\limits_{l = 1}^L { 2\mu Re \left\{ {Tr \left( {\mathbf{F}_l^{(n)H}\left( {{\mathbf{Z}_p} - {\mathbf{X}_p}} \right){\mathbf{F}_l}} \right)} \right\}} \\
\end{split} \;.
\end{equation}
The dual function can be obtained by solving the following problem
\begin{equation}\label{eq_scene1_dual_func}
g\left( \mu  \right) \buildrel \Delta \over = \mathop {\min} \limits_\mathbf{F} L\left( {\mathbf{F},\mu } \right) \;,
\end{equation}
and the dual problem is given by
\begin{equation}\label{eq_scene1_dual_pro}
\begin{split}
\mathop{\max}\limits_\mu \;& g \left( \mu \right) \\
s.t. \;& \mu \ge 0 \\
\end{split} \;.
\end{equation}
By setting the first-order derivative of $L\left( {\mathbf{F},\mu } \right)$ w.r.t. $\mathbf{F}$ to zero matrix, we can obtain the optimal
solution $\mathbf{\hat F}( \mu ) = \left\{ \mathbf{\hat F}_l( \mu ), \; \forall l \in {\cal L} \right\}$ as follows:
\begin{equation}\label{eq_scene1_solution}
{{\mathbf{\hat F}}_l}\left( \mu  \right) = {{\left( \mathbf{X}_0 + \mu \mathbf{Z}_p \right)}^\dag }\left( {\mathbf{Y}_l^H + \mu \left( {{\mathbf{Z}_p} - {\mathbf{X}_p}} \right) \mathbf{F}_l^{(n)}} \right)
\end{equation}
where pseudo inverse is adopted due to the fact that the matrix $\mathbf{X}_0 + \mu \mathbf{Z}_p$ is not full rank when $\mathbf{X}_0$ is not full rank
and $\mu = 0$. The value of $\mu$ should be chosen such that the complementary slackness condition for constraint (\ref{eq_scene1_inter}) is satisfied, namely,
\begin{equation}\label{eq_scene1_slack_cond}
\mu \left( \sum\limits_{l = 1}^L {Tr \left( {\mathbf{F}_l^H (\mu){\mathbf{Z}_p} {\mathbf{F}_l}(\mu)} \right)} \right. - \left. \sum\limits_{l = 1}^L { 2Re \left\{ {Tr \left( {\mathbf{F}_l^{(n)H}\left( {{\mathbf{Z}_p} - {\mathbf{X}_p}} \right){\mathbf{F}_l}(\mu)} \right)} \right\}} - {{\tilde \Gamma }_p} \right) = 0.
\end{equation}
Hence, if the following condition holds
\begin{equation}\label{eq_scene1_cons_cond}
\sum\limits_{l = 1}^L {Tr \left( {\mathbf{F}_l^H (0){\mathbf{Z}_p} {\mathbf{F}_l}(0)} \right)} - \sum\limits_{l = 1}^L { 2Re \left\{ {Tr \left( {\mathbf{F}_l^{(n)H}\left( {{\mathbf{Z}_p} - {\mathbf{X}_p}} \right){\mathbf{F}_l}(0)} \right)} \right\}} \le {{\tilde \Gamma }_p},
\end{equation}
the optimal solution to problem (\ref{eq_scene1_appro_pro}) is given by $\mathbf{\hat F} ( \mu ) |_{\mu  = 0}$. Otherwise, we need to find $\mu$ which satisfies the following equation:
\begin{equation}\label{eq_scene1_inter_power}
J(\mu) = \sum\limits_{l = 1}^L { Tr \left( {\mathbf{F}_l^H (\mu){\mathbf{Z}_p} {\mathbf{F}_l}(\mu)} \right)} - \sum\limits_{l = 1}^L { 2Re \left\{ {Tr \left( {\mathbf{F}_l^{(n)H}\left( {{\mathbf{Z}_p} - {\mathbf{X}_p}} \right){\mathbf{F}_l}(\mu)} \right)} \right\}} = {{\tilde \Gamma }_p}.
\end{equation}
For which, the following proposition is introduced.

\textbf{Proposition 2:} $J(\mu)$ is a monotonically non-increasing function of $\mu$.

\emph{Proof:} Please see the Appendix \ref{appendix_2}. $\hfill\blacksquare$

Based on Proposition 2, the bisection search method can be used to find the solution of equation (\ref{eq_scene1_inter_power}) and the algorithm is shown as below.
\begin{table}[H]
    \centering
    \begin{tabular}{l}
        \hline
        \textbf{Algorithm 4:} Bisection Search Method for (\ref{eq_scene1_inter_power}) \\
        \hline
        S1: Initialize $\varepsilon  > 0$, $n = 0$ and the bounds $\mu_l^{(0)}$ and $\mu_u^{(0)}$ of $\mu$; \\
        S2: Let ${\mu ^{(n)}} = (\mu _l^{(n)} + \mu _u^{(n)})/2$, calculate $\mathbf{\hat F}\left( \mu _u^{(n)}\right)$ and $J\left( {{\mu ^{(n)}}} \right)$ \\
        \quad according to (\ref{eq_scene1_solution}) and (\ref{eq_scene1_inter_power});\\
        S3: If $J\left( {{\mu ^{(n)}}} \right) \ge {\tilde \Gamma _p}$, set $\mu _l^{(n + 1)} = {\mu ^{(n)}}$ and $\mu _u^{(n + 1)} = \mu _u^{(n)}$; \\
        \quad otherwise set $\mu _l^{(n + 1)} = \mu _l^{(n)}$ and $\mu _u^{(n + 1)} = {\mu ^{(n)}}$;\\
        \quad Let $n = n+1$;\\
        S4: If $\left| {\mu _u^{(n)} - \mu _l^{(n)}} \right| > \varepsilon$, go back to S2; else, set $\hat \mu  = {\mu ^{(n)}}$;\\
        S5: Output $\hat{\mu}$;\\
        \hline
    \end{tabular}
    \label{algor_4}
\end{table}
In each iteration of Algorithm 4, we need to calculate $\mathbf{\hat F}\left( \mu\right)$ in (\ref{eq_scene1_solution}), which involves the calculation of ${\left( \mathbf{X}_0 + \mu \mathbf{Z}_p \right)}^\dag$ with a complexity of $O\left( {N_{SA}^3} \right)$. If the total number of iterations is $T$, the total complexity to calculate ${\left( \mathbf{X}_0 + \mu \mathbf{Z}_p \right)}^\dag$ is $O\left( T{N_{SA}^3} \right)$, which may be excessive. Here, we provide one method to reduce the computational complexity. Specifically, as $\mathbf{X}_0$ is a semi-definite positive matrix, it can be decomposed as $\mathbf{X}_0 = \mathbf{Q\Lambda }{\mathbf{Q}^H}$ by using the eigenvalue decomposition, where $\mathbf{Q}{\mathbf{Q}^H} = {\mathbf{Q}^H}\mathbf{Q} = \mathbf{I}$ and $\mathbf{\Lambda}$ is a diagonal matrix with non-negative diagonal elements. Then, we have ${\left( {{\mathbf{X}_0} + \mu {\mathbf{Z}_p}} \right)^\dag } = \mathbf{Q}{\left( {\mathbf{\Lambda } + \mu {\lambda _p}\mathbf{I}} \right)^\dag }{\mathbf{Q}^H}$ since ${\mathbf{Z}_p} = {\lambda _p}\mathbf{I}$. Hence, in each iteration, we only need to calculate the product of matrices, which has much lower complexity than calculating the inverse of matrices with the same dimension.

\underline{\emph{Case 2}}: Assuming the power constraint (28 C1) is an active constraint at $\mathbf{\hat F}$, problem (\ref{eq_appro_pro}) can be transformed into the following problem,
\begin{equation}\label{eq_scene2_appro_pro}
\begin{split}
\mathop{\min}\limits_\mathbf{F} \;& \sum\limits_{l = 1}^L {Tr \left( {\mathbf{F}_l^H{\mathbf{X}_0}{\mathbf{F}_l}} \right)} - \sum\limits_{l = 1}^L {2Re\left\{ {Tr \left( {\mathbf{Y}_l^H\mathbf{F}_l^H} \right)} \right\}}  \\
s.t. \;& C1:\sum\limits_{l = 1}^L { Tr \left( {\mathbf{F}_l^H{\mathbf{F}_l}} \right)} \le {P_{\max }} \\
&C2:\sum\limits_{l = 1}^L { 2Re \left\{ {Tr \left( {\mathbf{F}_l^{(n)H}\left( {{\mathbf{Z}_p} - {\mathbf{X}_p}} \right){\mathbf{F}_l}} \right)} \right\}} \ge {{\mathord{\buildrel{\lower3pt\hbox{$\scriptscriptstyle\frown$}}
\over \Gamma } }_p} \\
\end{split} \;.
\end{equation}
Herein, ${\mathord{\buildrel{\lower3pt\hbox{$\scriptscriptstyle\frown$}} \over \Gamma } _p} = {\lambda _p}{P_{\max }} - {\tilde \Gamma _p}$. By using Lagrangian dual decomposition method and introducing the Lagrange multiplier $\lambda$ associated with the power constraint, the partial Lagrangian function for problem (\ref{eq_scene2_appro_pro}) can be derived as follows
\begin{equation}\label{eq_scene2_part_lagrange_func}
L\left( {\mathbf{F},\lambda } \right) = \sum\limits_{l = 1}^L {{\rm{Tr}}\left( {\mathbf{F}_l^H{\mathbf{X}_0}{\mathbf{F}_l}} \right)}  - \sum\limits_{l = 1}^L {{\rm{2Re}}\left\{ {{\rm{Tr}}\left( {\mathbf{Y}_l^H\mathbf{F}_l^H} \right)} \right\}} + \lambda \sum\limits_{l = 1}^L { Tr \left( {\mathbf{F}_l^H{\mathbf{F}_l}} \right)} - \lambda {P_{\max }}.
\end{equation}
The dual function can be obtained by solving the following problem
\begin{equation}\label{eq_scene2_part_lagrange_func1}
\begin{split}
g\left( \lambda  \right) \;& \buildrel \Delta \over = \mathop {\min}\limits_\mathbf{F} L\left( {\mathbf{F},\lambda } \right)\\
s.t. \;&(40 \; C2)\\
\end{split}\;,
\end{equation}
and the dual problem is given by
\begin{equation}\label{eq_scene2_part_dual_pro}
\begin{split}
\mathop{\max}\limits_\lambda \;& g\left( \lambda \right) \\
s.t. \;& \lambda \ge 0 \\
\end{split} \;.
\end{equation}
In order to solve the dual problem (\ref{eq_scene2_part_dual_pro}), we need to derive the expression of dual function $g(\lambda)$ by solving problem (\ref{eq_scene2_part_lagrange_func1}) with given $\lambda$. By introducing dual variable $\mu \ge 0$ associated with the interference constraint (30), the Lagrangian function for problem (\ref{eq_scene2_part_lagrange_func1}) is given by
\begin{equation}\label{eq_scene2_lagrange_func}
\begin{split}
L\left( {\mathbf{F},\mu } \right) \;& = \sum\limits_{l = 1}^L {Tr \left( {\mathbf{F}_l^H\left( {{\mathbf{X}_0} + \lambda \mathbf{I}} \right){\mathbf{F}_l}} \right)} - \sum\limits_{l = 1}^L {2Re \left\{ {Tr \left( {\mathbf{Y}_l^H\mathbf{F}_l^H} \right)} \right\}}  - \lambda {P_{\max }} + \mu {{\mathord{ \buildrel{ \lower3pt \hbox{$\scriptscriptstyle\frown$}} \over \Gamma } }_p}\\
\;& - \sum\limits_{l = 1}^L { 2 \mu Re \left\{ {Tr \left( {\mathbf{F}_l^{(n)H}\left( {{\mathbf{Z}_p} - {\mathbf{X}_p}} \right){\mathbf{F}_l}} \right)} \right\}} \\
\end{split}
\end{equation}
By setting the first-order derivative of $L\left( {\mathbf{F},\mu } \right)$ w.r.t. $\mathbf{F}$ to zero matrix, we can obtain the optimal
solution $\mathbf{\hat F}( \mu ) = \left\{ \mathbf{\hat F}_l( \mu ), \; \forall l \in {\cal L} \right\}$ as follows
\begin{equation}\label{eq_scene2_solution}
{\mathbf{\hat F}_l} (\mu) = {\left( {{\mathbf{X}_0} + \lambda \mathbf{I}} \right)^\dag }\left( {\mathbf{Y}_l^H + \mu \left( {{\mathbf{Z}_p} - {\mathbf{X}_p}} \right)\mathbf{F}_l^{(n)}} \right) \;.
\end{equation}
Herein, the value of $\mu$ should be chosen such that the complementary slackness condition for constraint (\ref{eq_scene2_appro_pro} C2) is satisfied
\begin{equation}\label{eq_scene2_slack_cond}
\mu \left( \sum\limits_{l = 1}^L {2 Re \left\{ {Tr \left( {\mathbf{F}_l^{(n)H}\left( {{\mathbf{Z}_p} - {\mathbf{X}_p}} \right){{\mathbf{\hat F}}_l}(\mu)} \right)} \right\}} - {{\mathord{\buildrel{\lower3pt\hbox{$\scriptscriptstyle\frown$}} \over \Gamma } }_p} \right) = 0.
\end{equation}
Hence, if the following condition holds
\begin{equation}\label{eq_scene2_cons_cond}
\sum\limits_{l = 1}^L {2 Re \left\{ {Tr \left( {\mathbf{F}_l^{(n)H}\left( {{\mathbf{Z}_p} - {\mathbf{X}_p}} \right){{\mathbf{\hat F}}_l}(\mu){|_{\mu = 0}}} \right)} \right\}} \ge {\mathord{\buildrel{\lower3pt\hbox{$\scriptscriptstyle\frown$}} \over \Gamma } _p} \;,
\end{equation}
the optimal solution to (\ref{eq_scene2_part_lagrange_func1}) is given by $\mathbf{\hat F} ( \mu ) |_{\mu  = 0}$. Otherwise, the optimal $\mu$ is given by
\begin{equation}\label{eq_scene2_mu}
\mu = \frac{{{{\mathord{\buildrel{\lower3pt\hbox{$\scriptscriptstyle\frown$}} \over \Gamma } }_p} - \sum\limits_{l = 1}^L{ 2 Re \left\{{Tr \left( {{ \mathbf{F}_l^{(n)H} \left( {{\mathbf{Z}_p} - {\mathbf{X}_p}} \right){{\left( {{\mathbf{X}_0} + \lambda \mathbf{I}} \right)}^\dag }\mathbf{Y}_l^H} } \right)} \right\}}}} {{\sum\limits_{l = 1}^L {2 Tr\left( {\mathbf{F}_l^{(n)H}\left( {{\mathbf{Z}_p} - {\mathbf{X}_p}} \right){{\left( {{\mathbf{X}_0} + \lambda \mathbf{I}} \right)}^\dag }\left( {{\mathbf{Z}_p} - {\mathbf{X}_p}} \right)\mathbf{F}_l^{(n)}} \right)}}} \;.
\end{equation}
With dual function, we start to solve the dual problem (\ref{eq_scene2_part_dual_pro}) to find the optimal $\lambda$. Given $\lambda$, denote the optimal solution of the problem (\ref{eq_scene2_part_lagrange_func1}) as $\mathbf{\hat F}\left( \lambda \right)$. The value of $\lambda$ should be chosen such that
the complementary slackness condition for power constraint is satisfied
\begin{equation}\label{eq_scene2_slack_cond2}
\lambda \left( \sum\limits_{l = 1}^L { Tr \left( \mathbf{\hat F}_l^H (\lambda){\mathbf{\hat F}_l} (\lambda) \right)} - P_{\max} \right) = 0.
\end{equation}
If the following condition holds
\begin{equation}\label{eq_scene2_cons_cond2}
\sum\limits_{l = 1}^L {Tr \left( {\mathbf{\hat F}_l^H (\lambda)|_{\lambda = 0}{\mathbf{\hat F}_l} (\lambda)|_{\lambda = 0}} \right)} \le P_{\max},
\end{equation}
the optimal solution is given by $\mathbf{\hat F}\left( \lambda  \right){|_{\lambda  = 0}}$. Otherwise, we need to find $\lambda$ such that the following equation holds:
\begin{equation}\label{eq_scene2_max_power}
P(\lambda) = \sum\limits_{l = 1}^L {Tr \left( {\mathbf{\hat F}_l^H (\lambda){\mathbf{\hat F}_l}(\lambda) } \right)} = P_{\max}
\end{equation}
For which, the following proposition is introduced.

\textbf{Proposition 3:} $P(\lambda)$ is a monotonically non-increasing function of $\lambda$.

\emph{Proof:} The proof is similar to than for Proposition 2 and thus it is omitted herein. $\hfill\blacksquare$

Based on Proposition 3, the bisection search method can be used to find the solution of equation (\ref{eq_scene2_max_power}) and we formulate the algorithm shown below, i.e., the Algorithm 5.

\begin{table}[H]
    \centering
    \begin{tabular}{l}
        \hline
        \textbf{Algorithm 5:} Bisection Search Method for (\ref{eq_scene2_max_power}) \\
        \hline
        S1: Initialize $\varepsilon  > 0$, $n = 0$ and the bounds $\lambda_l^{(0)}$ and $\lambda_u^{(0)}$ of $\lambda$; \\
        S2: Let ${\lambda ^{(n)}} = (\lambda _l^{(n)} + \lambda _u^{(n)})/2$, if the condition (\ref{eq_scene2_cons_cond}) is satisfied, \\
        \quad $\mu = 0$; otherwise, calculate $\mu$ according to (\ref{eq_scene2_mu});\\
        S3: Calculate $\mathbf{\hat F}\left( \lambda _u^{(n)}\right)$ and $P\left( {{\lambda ^{(n)}}} \right)$ according to (\ref{eq_scene2_solution}) and (\ref{eq_scene2_max_power});\\
        S3: If $P\left( {{\lambda ^{(n)}}} \right) \ge P_{\max}$, set $\lambda _l^{(n + 1)} = {\lambda ^{(n)}}$ and $\lambda _u^{(n + 1)} = \lambda _u^{(n)}$; \\
        \quad otherwise set $\lambda _l^{(n + 1)} = \lambda _l^{(n)}$ and $\lambda _u^{(n + 1)} = {\lambda ^{(n)}}$;\\
        \quad Let $n = n+1$;\\
        S4: If $\left| {\lambda _u^{(n)} - \lambda _l^{(n)}} \right| > \varepsilon$, go back to S2; else, set $\hat \lambda  = {\lambda ^{(n)}}$;\\
        S5: Output $\hat{\lambda}$;\\
        \hline
    \end{tabular}
    \label{algor_5}
\end{table}
Similarly, in each iteration of Algorithm 5, we need to calculate $\mathbf{\hat F}\left( \lambda \right)$ in (\ref{eq_scene2_solution}), which involves the calculation of ${\left( \mathbf{X}_0 + \lambda \mathbf{I}\right)}^\dag$. Herein, the same approach as Algorithm 4 can be adopted to reduce the complexity.

Based on the above discussion, the details of the SCA algorithm to solve OP7 is summarized as the following Algorithm 6.
\begin{table}[H]
    \centering
    \begin{tabular}{l}
        \hline
        \textbf{Algorithm 6:} SCA-based Algorithm to Solve OP7\\
        \hline
        S1: Initialize $\varepsilon >0$, $\mathbf{F} ^{(0)}$, $n = 0$, calculate the objective value of the OP7 \\
        \quad as $z \left( \mathbf{F} ^{(0)} \right)$;\\
        S2: Given $\mathbf{F}^{(0)}$, obtain $\mathbf{\hat F}^{(n)}$ by solving (\ref{eq_scene1_appro_pro}) using algorithm 4;\\
        S3: if $\mathbf{\hat F}^{(n)}$ satisfies the constraint (\ref{eq_OP7} C1), go to S5;\\
        S4: Given $\mathbf{F}^{(0)}$, obtain $\mathbf{\hat F}^{(n)}$ by solving (\ref{eq_scene2_appro_pro}) using algorithm 5;\\
        S5: If $\left| {z\left( \mathbf{\hat F}^{(n)} \right) - z\left( \mathbf{F}^{(n)} \right)} \right| > \varepsilon$, set $z\left( \mathbf{F}^{(n+1)} \right) = z\left( \mathbf{\hat F}^{(n)} \right)$, \\
        \quad $\mathbf{F}^{(n+1)} = \mathbf{\hat F}^{(n)}$, $n=n+1$, go back to S2; else set $\mathbf{\hat F}=\mathbf{\hat F}^{(n)}$;\\
        S6: Output $\mathbf{\hat{F}}$;\\
        \hline
    \end{tabular}
    \label{algor_6}
\end{table}

Also, we have the convergence conclusion for the Algorithm 6 as below.

\textbf{Proposition 4:} The sequence generated by Algorithm 6, i.e., $\left\{ \mathbf{\hat F}^{(n)},n = 0,1,2,... \right\}$ converges to the KKT optimum point of OP7.

\emph{Proof:} Please see the Appendix \ref{appendix_3}. $\hfill\blacksquare$

Now, we briefly analyze the complexity of Algorithm 6. Firstly, assuming the number of iterations for Algorithm 6 is $T$, the number of iterations for Algorithm 4 and Algorithm 5 to converge are given by $\log _2 \left( \frac{{{\mu _u} - {\mu _l}}}{\varepsilon } \right)$ and $\log _2\left( \frac{{{\lambda _u} - {\lambda _l}}} {\varepsilon } \right)$, respectively. Note that, the main complexity lies in calculating $\mathbf{F}$ in each iteration of Algorithm 6. Taking advantage of the structures of $\mathbf{X}_0 + \mu \mathbf{Z}_p$ and $\mathbf{X}_0 + \lambda \mathbf{I}$, the computation of ${\left( \mathbf{X}_0 + \mu \mathbf{Z}_p \right)}^\dag$ and ${\left( {{\mathbf{X}_0} + \lambda \mathbf{I}} \right)^\dag }$ can be simplified as the product of matrices by the eigenvalue decomposition of $\mathbf{X}_0$ before entering algorithm, whose complexities is $O\left( N_{SA}^3 \right)$. Assuming the complexities of calculating $\mathbf{F}$ in each iteration of Algorithm 4 and Algorithm 5 are denoted by $n_1$ and $n_2$, therefore, the total complexity of Algorithm 6 is given by $O\left( {N_{SA}^3 + TL\left( {{n_1}{{\log }_2}\left( {\frac{{{\mu _u} - {\mu _l}}}{\varepsilon }} \right) + {n_2}{{\log }_2}\left( {\frac{{{\lambda _u} - {\lambda _l}}}{\varepsilon }} \right)} \right)} \right)$.

\subsection{Optimize the IRS reflecting coefficients}
In this subsection, we focus on optimizing the reflecting coefficients at the IRS while fixing the other parameters. Based on the problem (\ref{eq_IRS_reform4}), the reflecting coefficients optimization problem in the simplified model is given by the following OP8,
\begin{equation}\label{eq_OP8}
\begin{split}
\mathop{\min}\limits_\mathbf{\Theta} \;& Tr\left( {{\mathbf{B}_0}\mathbf{\Theta C}{\mathbf{\Theta }^H}} \right) + 2Re\left\{ {Tr \left( {\mathbf{D}_0^H {\mathbf{\Theta}^H} } \right)} \right\} \\
s.t. \;& Tr \left( {{\mathbf{B}_p} \mathbf{\Theta C} {\mathbf{\Theta }^H}} \right) + 2Re\left\{ {Tr \left( {\mathbf{D}_p^H {\mathbf{\Theta}^H}} \right)} \right\} \le {{\tilde \Gamma }_p} \\
\;& \left| {{\mathbf{\Theta}_{mm}}} \right| = 1, \; \forall m \in {\cal M} \\
\end{split} \;.
\end{equation}
Therein, $\mathbf{B}_p = \mathbf{H}_{rp}^H \mathbf{H}_{rp} \succcurlyeq \mathbf{0}$, $\mathbf{D}_p = \mathbf{H}_{sr} \mathbf{Q}_s \mathbf{H}_{sp}^H \mathbf{H}_{rp}$ and ${\tilde \Gamma _p} = {\Gamma _p} - Tr \left( \mathbf{H}_{sp} \mathbf{Q}_s \mathbf{H}_{sp}^H \right)$. Same as the last section, OP8 can be further transformed to
\begin{equation}\label{eq_OP8_reform1}
\begin{split}
\mathop{\min}\limits_{\boldsymbol{\theta}} \;& {{\boldsymbol{\theta}}^H}{\mathbf{\Upsilon}_0}{\boldsymbol{\theta }} + 2Re\left\{ {{{\boldsymbol{\theta}}^H} {\boldsymbol{d}}_0^*} \right\} \\
s.t. \;& {{\boldsymbol{\theta}}^H}{\mathbf{\Upsilon}_p}{\boldsymbol{\theta}} + 2Re \left\{ {{{\boldsymbol{\theta}}^H}{\boldsymbol{d}}_p^*} \right\} \le {{\tilde \Gamma }_p} \\
\;& \left| {{{\boldsymbol{\theta }}_m}} \right| = 1,\;\forall m \in {\cal M} \\
\end{split} \;.
\end{equation}

Herein, $\mathbf{\Upsilon}_p = \mathbf{B}_p \odot \mathbf{C}^T \succcurlyeq \mathbf{0}$ and $\boldsymbol{d}_p = \left[ {{\left[ {{\mathbf{D}_p}} \right]}_{1,1}},{{\left[ {{\mathbf{D}_p}} \right]}_{2,2}},...,{{\left[ {{\mathbf{D}_p}} \right]}_{M,M}} \right]^T$. Note that, the problem (\ref{eq_OP8_reform1}) could not be solved directly since the non-convexity of the uni-modular constraint, hence, SCA approach is adopted again. For which, we introduce the following proposition.

\textbf{Proposition 5:} Let $f_0\left(\boldsymbol{\theta }\right)={{\boldsymbol{\theta }}^H}{\mathbf{\Upsilon} _0}{\boldsymbol{\theta }}$, ${\mathbf{\Upsilon} _0} \succcurlyeq \mathbf{0}$, $\mathbf{X}_0 = {\lambda _{0,\max }}{\mathbf{I}_{M \times M}}$ and $\lambda _{0,\max }$ denotes the maximum eigenvalue of the ${\mathbf{\Upsilon} _0}$. Hence, for $\forall \boldsymbol{\theta }$, given $\boldsymbol{\theta }^{(n)}$, there exists
\begin{equation}\label{eq_lamma5_1}
{\tilde f}_0\left(\boldsymbol{\theta }|\boldsymbol{\theta }^{(n)}\right)= {{\boldsymbol{\theta }}^H}{\mathbf{X}_0}{\boldsymbol{\theta }} - 2 Re \left\{ {{{\boldsymbol{\theta }}^H}\left( {{\mathbf{X}_0} - {\mathbf{\Upsilon}_0}} \right){{\boldsymbol{\theta }}^{(n)}}} \right\}
+ {\left( {{{\boldsymbol{\theta }}^{(n)}}} \right)^H}\left( {{\mathbf{X}_0} - {\mathbf{\Upsilon}_0}} \right){{\boldsymbol{\theta }}^{(n)}}
\end{equation}
which satisfies the following three conditions:

1) ${\tilde f_0}\left( {{\boldsymbol{\theta }^{(n)}}|{\boldsymbol{\theta }^{(n)}}} \right) = {f_0}\left( {{\boldsymbol{\theta }^{(n)}}} \right)$,

2) ${\left. {{\nabla _{{\boldsymbol{\theta }^*}}}{{\tilde f}_0}\left( {\boldsymbol{\theta }|{\boldsymbol{\theta }^{(n)}}} \right)} \right|_{\boldsymbol{\theta } = {\boldsymbol{\theta }^{(n)}}}} = {\left. {{\nabla _{{\boldsymbol{\theta }^*}}}{f_0}\left( \boldsymbol{\theta } \right)} \right|_{\boldsymbol{\theta } = {\boldsymbol{\theta }^{(n)}}}}$,

3) ${\tilde f_0}\left( {\boldsymbol{\theta }|{\boldsymbol{\theta }^{(n)}}} \right) \ge {f_0}\left( \boldsymbol{\theta } \right)$.

\emph{Proof:} The proof is similar to that for Proposition 1 and thus it is omitted herein. $\hfill\blacksquare$

Same as ${\tilde f}_0\left(\boldsymbol{\theta }|\boldsymbol{\theta }^{(n)}\right)$ in Proposition 5, let
\begin{equation}\label{eq_lamma5_1}
{\tilde f}_p\left(\boldsymbol{\theta }|\boldsymbol{\theta }^{(n)}\right) = {{\boldsymbol{\theta }}^H}{\mathbf{X}_p}{\boldsymbol{\theta }} - 2 Re \left\{ {{{\boldsymbol{\theta }}^H}\left( {{\mathbf{X}_p} - {\mathbf{\Upsilon}_p}} \right){{\boldsymbol{\theta }}^{(n)}}} \right\} + {\left( {{{\boldsymbol{\theta }}^{(n)}}} \right)^H}\left( {{\mathbf{X}_p} - {\mathbf{\Upsilon}_p}} \right){{\boldsymbol{\theta }}^{(n)}}.
\end{equation}
Given ${\boldsymbol{\theta }}^{(n)}$, we have
\begin{equation}\label{eq_OP8_appro0}
\begin{split}
\mathop{\max}\limits_{\boldsymbol{\theta }} \;& {{\tilde f}_0}\left( {{\boldsymbol{\theta }}|{{\boldsymbol{\theta }}^{(n)}}} \right) + 2Re\left\{ {{{\boldsymbol{\theta }}^H}{\boldsymbol{d}}_0^*} \right\} \\
s.t. \;& {{\tilde f}_p}\left( {{\boldsymbol{\theta }}|{{\boldsymbol{\theta }}^{(n)}}} \right) + 2Re\left\{ {{{\boldsymbol{\theta }}^H}{\boldsymbol{d}}_p^*} \right\} \le \tilde \Gamma _p \\
\;& \left| {{{\boldsymbol{\theta }}_m}} \right| = 1, \; \forall m \in {\cal M} \\
\end{split}
\end{equation}
Since ${{\boldsymbol{\theta }}^H} {\boldsymbol{\theta }} = M$, we have ${{\boldsymbol{\theta }}^H}{\mathbf{X}_0}{\boldsymbol{\theta }} = M{\lambda _{0,\max }}$ and ${{\boldsymbol{\theta }}^H}{\mathbf{X}_p}{\boldsymbol{\theta }} = M{\lambda _{p,\max }}$, which is a constant. By removing all constant terms,
the problem (\ref{eq_OP8_appro0}) can be rewritten as follows:
\begin{equation}\label{eq_OP8_appro1}
\begin{split}
\mathop{\max}\limits_{\boldsymbol{\theta }} \;& Re \left\{ {{{\boldsymbol{\theta }}^H}{\boldsymbol{q}}_0^{(n)}} \right\} \\
s.t. \;& C1: Re\left\{ {{{\boldsymbol{\theta }}^H}{\boldsymbol{q}}_p^{(n)}} \right\} \ge \tilde \Gamma _p^{(n)} \\
\;& C2: \left| {{{\boldsymbol{\theta }}_m}} \right| = 1, \; \forall m \in {\cal M} \\
\end{split} \;,
\end{equation}
where $\tilde \Gamma _p^{(n)} = \left[ {M{\lambda _{p,\max }} + {{\left( {{{\boldsymbol{\theta }}^{(n)}}} \right)}^H}\left( {{\mathbf{X}_p} - {\Upsilon _p}} \right){{\boldsymbol{\theta }}^{(n)}} - {{\tilde \Gamma }_p}} \right] \mathord{\left/
 {\vphantom {{\left[ {M{\lambda _{p,\max }} + {{\left( {{{\boldsymbol{\theta }}^{(n)}}} \right)}^H}\left( {{\mathbf{X}_p} - {\Upsilon _p}} \right){{\boldsymbol{\theta }}^{(n)}} - {{\tilde \Gamma }_p}} \right]} 2}} \right.
 \kern-\nulldelimiterspace} 2$, ${\boldsymbol{q}}_0^{(n)} = \left( {{\lambda _{0,\max }}{\mathbf{I}_{M \times M}} - \mathbf{\Upsilon} _0} \right){{\boldsymbol{\theta }}^{(n)}} - {\boldsymbol{d}}_0^*$ and ${\boldsymbol{q}}_p^{(n)} = \left( {{\lambda _{p,\max }}{\mathbf{I}_{M \times M}} - \mathbf{\Upsilon} _p} \right){{\boldsymbol{\theta }}^{(n)}} - {\boldsymbol{d}}_p^*$. The problem (\ref{eq_OP8_appro1}) could not be solved directly since the non-convexity of the uni-modular constraint. Therefore, a price mechanism is introduced to solve the problem (\ref{eq_OP8_appro1}) that can obtain the globally optimal solution. In specific, we consider the following problem by introducing $\alpha \ge 0$:
\begin{equation}\label{eq_OP8_appro2}
\begin{split}
\mathop{\max}\limits_{\boldsymbol{\theta }} \;& Re \left\{ {{{\boldsymbol{\theta }}^H}{\boldsymbol{q}}_0^{(n)}} \right\} + \alpha Re \left\{ {{{\boldsymbol{\theta }}^H}{\boldsymbol{q}}_p^{(n)}} \right\} \\
\;& \left| {{{\boldsymbol{\theta }}_m}} \right| = 1, \; \forall m \in {\cal M} \\
\end{split}
\end{equation}
For given $\alpha$, the globally optimal solution is given by
\begin{equation}\label{eq_OP8_solution}
{\boldsymbol{\theta}}(\alpha) = {e^{j\arg \left( {q_0^{(n)} + \alpha q_p^{(n)}} \right)}}
\end{equation}
Our objective is to find a $\alpha$ such that the following complementary slackness condition is satisfied:
\begin{equation}\label{eq_OP8_slack_cond}
\alpha \left( {g\left( \alpha  \right) - \tilde \Gamma _p^{(n)}} \right) = 0 \;.
\end{equation}
Herein, $g(\alpha) = Re \left\{ {{\boldsymbol{\theta }}{{(\alpha)}^H}{\boldsymbol{q}}_p^{(n)}} \right\}$. To solve this equation, we consider two cases: 1) $\alpha = 0$; 2) $\alpha > 0$;.

\underline{\emph{Case 1}}: Consider $\alpha = 0$, $\boldsymbol{\theta}(0) = e^{j\arg \left( \boldsymbol{q}_0^{(n)} \right)}$ needs to satisfy constraint (\ref{eq_OP8_appro1} C1). Otherwise, $\alpha > 0$.

\underline{\emph{Case 2}}: Consider $\alpha > 0$,  equation (\ref{eq_OP8_slack_cond}) holds only when $g(\alpha) = \tilde \Gamma _p^{(n)}$. To solve this equation, we first provide the following proposition.

\textbf{Proposition 6:} $g(\alpha)$ is a monotonically non-decreasing function of $\alpha$.

\emph{Proof:} The proof is similar to that for Proposition 2 and thus it is omitted herein. $\hfill\blacksquare$

Based on Proposition 6, the bisection search method can be adopted to find the solution of the equation (\ref{eq_OP8_slack_cond}) and the algorithm is provided in the following.
\begin{table}[H]
    \centering
    \begin{tabular}{l}
        \hline
        \textbf{Algorithm 7:} Bisection Search Method for (\ref{eq_OP8_slack_cond}) \\
        \hline
        S1: Calculate $g(0)$. If $g(0) \le \tilde \Gamma _p^{(n)}$, set $\hat \alpha  = 0$ and terminate. \\
        \quad Otherwise, go to S2.\\
        S2: Initialize $\varepsilon  > 0$, $n = 0$ and the bounds $\alpha_l^{(0)}$ and $\alpha_u^{(0)}$ of $\alpha$; \\
        S3: Set ${\alpha ^{(n)}} = (\alpha _l^{(n)} + \alpha _u^{(n)})/2$ and calculate $g\left( {{\alpha ^{(n)}}} \right)$; \\
        S4: If $g\left( \alpha ^{(n)} \right) \ge {\tilde \Gamma _p^{(n)}}$, set $\alpha _l^{(n + 1)} = {\alpha ^{(n)}}$ and $\alpha _u^{(n + 1)} = \alpha _u^{(n)}$; \\
        \quad otherwise set $\alpha _l^{(n + 1)} = \alpha _l^{(n)}$ and $\alpha _u^{(n + 1)} = {\alpha ^{(n)}}$;\\
        \quad Let $n = n+1$;\\
        S5: If $\left| {\alpha _u^{(n)} - \alpha _l^{(n)}} \right| > \varepsilon$, go back to S3; else, set $\hat \alpha  = {\alpha ^{(n)}}$;\\
        S6: Output $\hat{\alpha}$;\\
        \hline
    \end{tabular}
    \label{algor_7}
\end{table}

Although the problem (\ref{eq_OP8_appro1}) is a non-convex problem, in the following theorem, we prove that Algorithm 7 can obtain the globally optimal solution.

\textbf{Theorem 7:} Algorithm 7 can obtain the globally optimal solution of problem (\ref{eq_OP8_appro1}).

\emph{Proof:} The proof is similar to the proof of Theorem 2 in \cite{DBLP:journals/jsac/PanRWENWH20} and omitted for simplicity.

Based on the above, we now provide the details of solving OP8 in Algorithm 8.
\begin{table}[H]
    \centering
    \begin{tabular}{l}
        \hline
        \textbf{Algorithm 8:} SCA-based Algorithm To Solve The OP8\\
        \hline
        S1: Initialize $\boldsymbol{\theta }^{(0)}$, $\varepsilon >0$, $n = 0$, calculate the objective value of the OP8 \\
        \quad as $z \left( \boldsymbol{\theta } ^{(0)} \right)$;\\
        S2: Given $\boldsymbol{\theta }^{(0)}$, obtain $\boldsymbol{\hat \theta}^{(n)}$ by solving (\ref{eq_OP8_appro1}) using algorithm 7;\\
        S3: If $\left| {z\left( \boldsymbol{\hat \theta}^{(n)} \right) - z\left( \boldsymbol{\theta }^{(n)} \right)} \right| > \varepsilon$, set $z\left( \boldsymbol{\theta }^{(n+1)} \right) = z\left( \boldsymbol{\hat \theta}^{(n)} \right)$, \\
        \quad $\boldsymbol{\theta }^{(n+1)} = \boldsymbol{\hat \theta}^{(n)}$, $n=n+1$, go back to S2; else set $\boldsymbol{\hat \theta}=\boldsymbol{\hat \theta}^{(n)}$;\\
        S4: Output $\boldsymbol{\hat \theta}$;\\
        \hline
    \end{tabular}
    \label{algor_8}
\end{table}
In the following theorem, we prove that the sequence of $\left\{\boldsymbol{\hat \theta}^{(n)},n = 0,1,2,... \right\}$ generated by Algorithm 8 converges to the KKT optimal point of OP8.

\textbf{Theorem 8:} The sequences of the objective value produced by Algorithm 8 is guaranteed to converge, and the final solution satisfies the KKT point of the OP8.

\emph{Proof:} The proof is similar to the proof of Theorem 3 in \cite{DBLP:journals/jsac/PanRWENWH20} and omitted for simplicity.

Now, we further analyze the complexity of Algorithm 8. Firstly, we assume that the number of iterations for Algorithm 8 is $T$. Note that, the main complexity of the Algorithm 8 lies in calculating the maximum eigenvalue of $\mathbf{\Upsilon}_0$ and $\mathbf{\Upsilon}_p$ and the bisection search in the Algorithm 7. Herein, the maximum eigenvalue of $\mathbf{\Upsilon}_0$ and $\mathbf{\Upsilon}_p$, whose complexities are $O\left( {{M^3}} \right)$, only need to be calculated once before entering algorithm. Moreover, the number of iterations for Algorithm 7 to converge is characterized by ${\log _2}\left( {\frac{{{\alpha _u} - {\alpha _l}}}{\varepsilon }} \right)$. By assuming that the complexities of each iteration of the bisection search in the Algorithm 7 is denoted by $n_1$, the total complexity of Algorithm 8 is given by $O\left( {{M^3} + T{n_1}{{\log }_2}\left( {\frac{{{\alpha _u} - {\alpha _l}}}{\varepsilon }} \right)} \right)$.

\subsection{Overall Algorithm}
In this subsection, the overall algorithm based on alternating optimization for OP5 is provided and summarized as the following Algorithm 9.
\begin{table}[H]
    \centering
    \begin{tabular}{l}
        \hline
        \textbf{Algorithm 9:} AO-based Algorithm to Solve OP5\\
        \hline
        S1: Initialize $\mathbf{F} ^{(0)}$, $\boldsymbol{\theta}^{(0)}$, $\varepsilon >0$, $n = 0$, calculate the WSR of all SUs as \\ \quad $R\left( \mathbf{F} ^{(0)}, \mathbf{\Theta} ^{(0)} \right)$;\\
        S2: Given $\mathbf{F} ^{(n)}$ and $\boldsymbol{\theta} ^{(n)}$, obtain $\mathbf{\hat U}^{(n)}$ and $\mathbf{\hat W}^{(n)}$ according to (\ref{eq_U_l}) and \\
        \quad (\ref{eq_W_l});\\
        S3: Given $\mathbf{\hat U}^{(n)}$, $\mathbf{\hat W}^{(n)}$ and $\boldsymbol{\theta}^{(n)}$, obtain $\mathbf{\hat F}^{(n)}$ by solving the OP7 using \\
        \quad algorithm 6;\\
        S4: Given $\mathbf{\hat U}^{(n)}$, $\mathbf{\hat W}^{(n)}$ and $\mathbf{\hat F}^{(n)}$, obtain $\boldsymbol{\hat \theta}^{(n)}$ by solving the OP8 using \\
        \quad algorithm 8;\\
        S5: If $\left| {R\left( {{\mathbf{\hat F}^{(n)}},{\mathbf{\hat \Theta }^{(n)}}} \right) - R\left( {{\mathbf{F}^{(n)}} , {\mathbf{\Theta }^{(n)}}} \right)} \right| > \varepsilon$, set $\mathbf{F}^{(n+1)} = \mathbf{\hat F}^{(n+1)}$\\
        \quad $\boldsymbol{\theta}^{(n+1)}=\boldsymbol{\hat{\theta}}^{(n)}$, ${R\left( {{\mathbf{F}^{(n+1)}},{\mathbf{\Theta }^{(n+1)}}} \right) = R\left( {{\mathbf{\hat F}^{(n)}} , {\mathbf{\hat \Theta}^{(n)}}} \right)}$, \\
        \quad $n=n+1$, go back to S2; else set $\mathbf{\hat F}=\mathbf{\hat F}^{(n)}$, $\boldsymbol{\hat{\theta}}=\boldsymbol{\hat{\theta}}^{(n)}$;\\
        S6: Output $\mathbf{\hat{F}}$ and $\boldsymbol{\hat{\theta}}$;\\
        \hline
    \end{tabular}
    \label{algor_9}
\end{table}

\noindent Herein, $\mathbf{\hat U}^{(n)}$, $\mathbf{\hat W}^{(n)}$, $\mathbf{\hat F}^{(n)}$ and $\boldsymbol{\hat \theta}^{(n)}$ represent the stable solutions obtained by solving the subproblems in the $n$th iteration, and $R\left( \mathbf{F,\Theta } \right)$ denotes the WSR of all SUs. Since the original problem is bounded and the progress of the alternative optimization is monotonically non-decreasing, thus the above algorithm is surely convergent. Furthermore, the complexity of Algorithm 9 mainly depends on Algorithm 7 in Step 3 and Algorithm 8 in Step 4, whose complexities have been explained in the above, which will be omitted here due to space limits.

\section{Simulation Analysis}\label{section_5}
In this section, the performances of the proposed algorithms are evaluated by numerical simulations. Corresponding to the general scenario with multiple PUs in Section \ref{section_3} and the special scenario with only one PU in Section \ref{section_4}, simulation results are given by two subsections.

\subsection{General Scenario With Multiple PUs}
In this subsection, simulation results regarding the general scenario are provided. In the IRS-MIMO-CR system under our consideration, the SAP and the IRS are located at $(0, 0)$ and $(30, 5)$ in meter ($m$) in a two-dimensional plane, respectively. In addition, there are three PUs and three SUs and they are uniformly and randomly scattered in a circle with radius 2 m and centered at $(50, 0)$ and $(30, 0)$, respectively. The other system parameters used in the simulations are referred to \cite{DBLP:journals/access/JiangZWFJ20} and \cite{DBLP:journals/jsac/ZhangZLLDL20}, that is, we set the antenna numbers of the SAP, PUs and SUs as $N_{SA} = 4$ and $N_{PU} = N_{SU} = 2$, respectively. The number of data streams for each SU is set as $d = 2$. The noise power at all PUs and SUs is set as $\sigma _R^2=\sigma _E^2=-40dBm, \; \forall k \in {\cal K},l \in {\cal L}$ and the maximum total transmitted power is set as $P_{\max}=5W$. The maximum interference at all PUs is set as $I{T_k} = 2 \times {10^{-4}}W, \; \forall k \in {\cal K}$. Without loss of generality, all the channels are modeled as \cite{DBLP:journals/twc/WuZ19}
\begin{equation*}
\mathbf{H}=\sqrt{{\beta }/{\kappa +1}\;}\left( \sqrt{\kappa }{{\mathbf{H}}^{LoS}}+{{\mathbf{H}}^{NLoS}} \right),
\end{equation*}

\noindent where $\kappa$ is the Rician factor, while $\mathbf{H}^{LOS}$ and $\mathbf{H}^{NLOS}$ represent the deterministic line-of-sight (LoS) and Rayleigh fading/non-LoS (NLoS) components, respectively. $\beta$ represents the path loss, and is given by $\beta ={{\beta }_{0}}-10\alpha {{\log }_{10}}\left( {d}/{{{d}_{0}}}\; \right)$. Herein, $\beta _0$ denotes the path loss at the reference distance $d_0=1m$, $\alpha$ and $d$ represent the path loss exponent and the distance between the corresponding nodes, respectively. Assuming that the location of IRS can be carefully selected, the channels from IRS to PUs and SUs have LoS component and experience Rayleigh fading, simultaneously. However, the channels from SAP to PUs, SUs and IRS, only experience Rayleigh fading. Hence, the Rician factors are set as ${{\kappa}_{rk}}={{\kappa}_{rl}}=1$ and ${{\kappa }_{sk}}={{\kappa }_{sl}}={{\kappa}_{sr}}=0$. In addition, path loss exponents of all channels are set as ${{\alpha}_{sk}}={{\alpha}_{sl}}={{\alpha}_{sr}}={{\alpha}_{rk}}={{\alpha}_{rl}}=2$.

Furthermore, in order to better understand the positive effects of the IRS in improving the performance for the IRS-MIMO-CR system and the performance gain of the proposed algorithms, some benchmark schemes are introduced in simulations for performance comparison and analysis. Thus, following three algorithms are evaluated in the simulations, i.e., `no-IRS', `fixed-IRS' and `AO'.

\textbf{\emph{no-IRS}:} That is, no IRS is used in the system and the WSR is obtained by directly optimizing OP1 under the conditions $\mathbf{G}_{sk}(\mathbf{\Theta}) = \mathbf{H}_{sk}$ and $\mathbf{G}_{sl}(\mathbf{\Theta}) = \mathbf{H}_{sl}$.

\textbf{\emph{fixed-IRS}:} That is, all reflecting coefficients of the IRS have same phases, which are set to zeros, namely, $arg\left( \boldsymbol{\theta}_m \right) = 0, m=1,..,M$.

\textbf{\emph{AO}:} It is our proposed AO-based algorithm, i.e., the Algorithm 3.

\begin{figure}[h]
    \centering
    \includegraphics[scale=0.6]{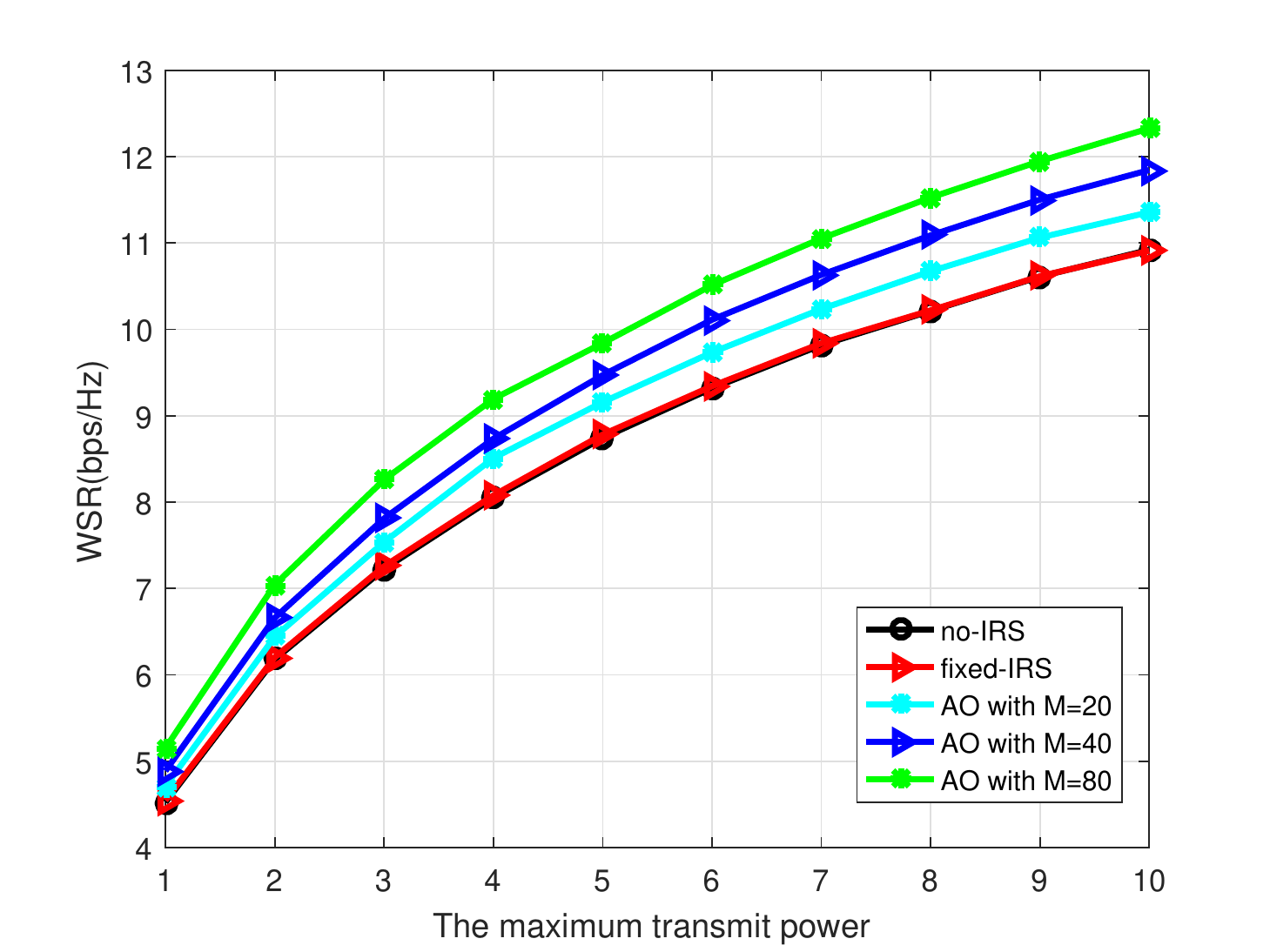}
    \caption{\quad WSR vs the maximum transmit power}
    \label{RateVsPower_M}
\end{figure}

At first, the WSR of different algorithms is evaluated by varying the available transmission power at the SAP, i.e., the $P_{max} \in [1,10]W$, and the result is shown in Fig. \ref{RateVsPower_M}. The `AO with $M = 20$', `AO with $M = 40$', and `AO with $M = 80$' shown in the figure are used to identify Algorithm 3 when the number of IRS reflection elements is 20, 40, and 80, respectively. Apparently, with the increase of the available transmission power at the SAP, the WSR for all these algorithms is increased. Moreover, algorithms `no-IRS' and `fixed-IRS' obtain the worst performance and they are exceedingly close. Introducing the IRS optimization, the performance of `AO with $M = 20$', `AO with $M = 40$' and `AO with $M = 80$' is improved clearly with the increase of the number of IRS reflection elements and better than that of `no-IRS' and `fixed-IRS'. It is worth mentioning that the performance gaps between the various mechanisms become larger with the increase of the transmission power at the SAP, which means that the introduction of the IRS reflection coefficient optimization brings more significant performance gains at the higher transmission power.

\begin{figure}[h]
    \centering
    \includegraphics[scale=0.6]{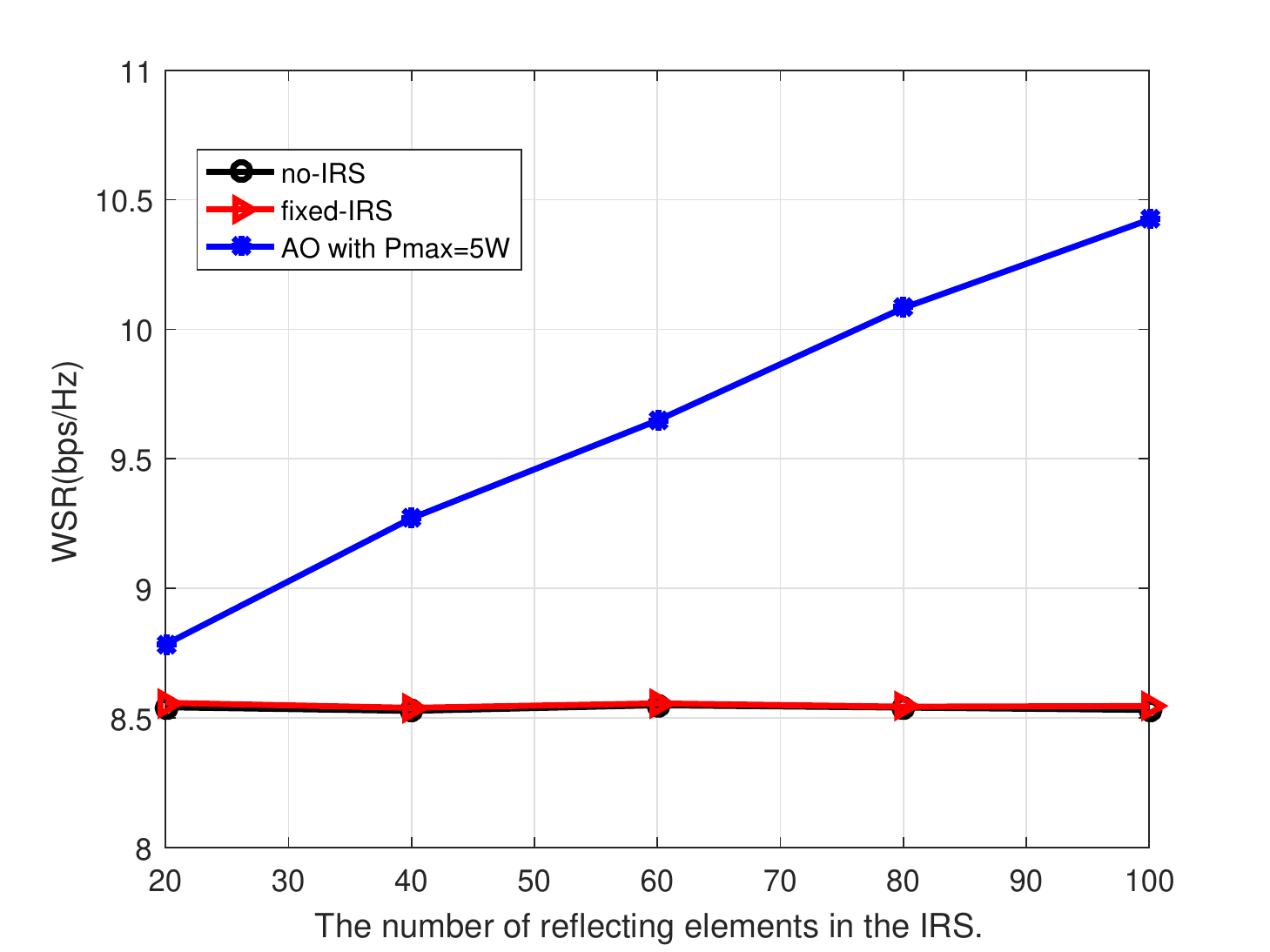}
    \caption{\quad WSR vs the number of reflecting elements}
    \label{RateVsIRS_M}
\end{figure}

Clearly, the performance of Algorithm 3 improves significantly as the number of IRS reflection units increases in Fig. \ref{RateVsPower_M}. Then, in order to study the impact of the number of IRS reflection elements, we use Fig. \ref{RateVsIRS_M} to show the trend of the WSR with the number of IRS reflection elements under three mechanisms. The maximum transmission power of the SAP is set as $P_{\max} = 5W$, and the number of IRS reflection elements varies from 20 to 100. The `AO with $P_{\max} = 5W$' shown in the figure is used to identify Algorithm 3 with $P_{\max} = 5W$. Note that, the curve corresponding to `AO with $P_{\max} = 5W$' shows a significant upward trend, and is significantly improved compared to the other two mechanisms, while the performance of `no-IRS' and `fixed-IRS' mechanisms is not affected by the number of the IRS elements in the system. The fact means that the wireless communication environment can be improved by increasing the number of IRS reflection elements appropriately.

\begin{figure}[tbp]
    \centering
    \includegraphics[scale=0.6]{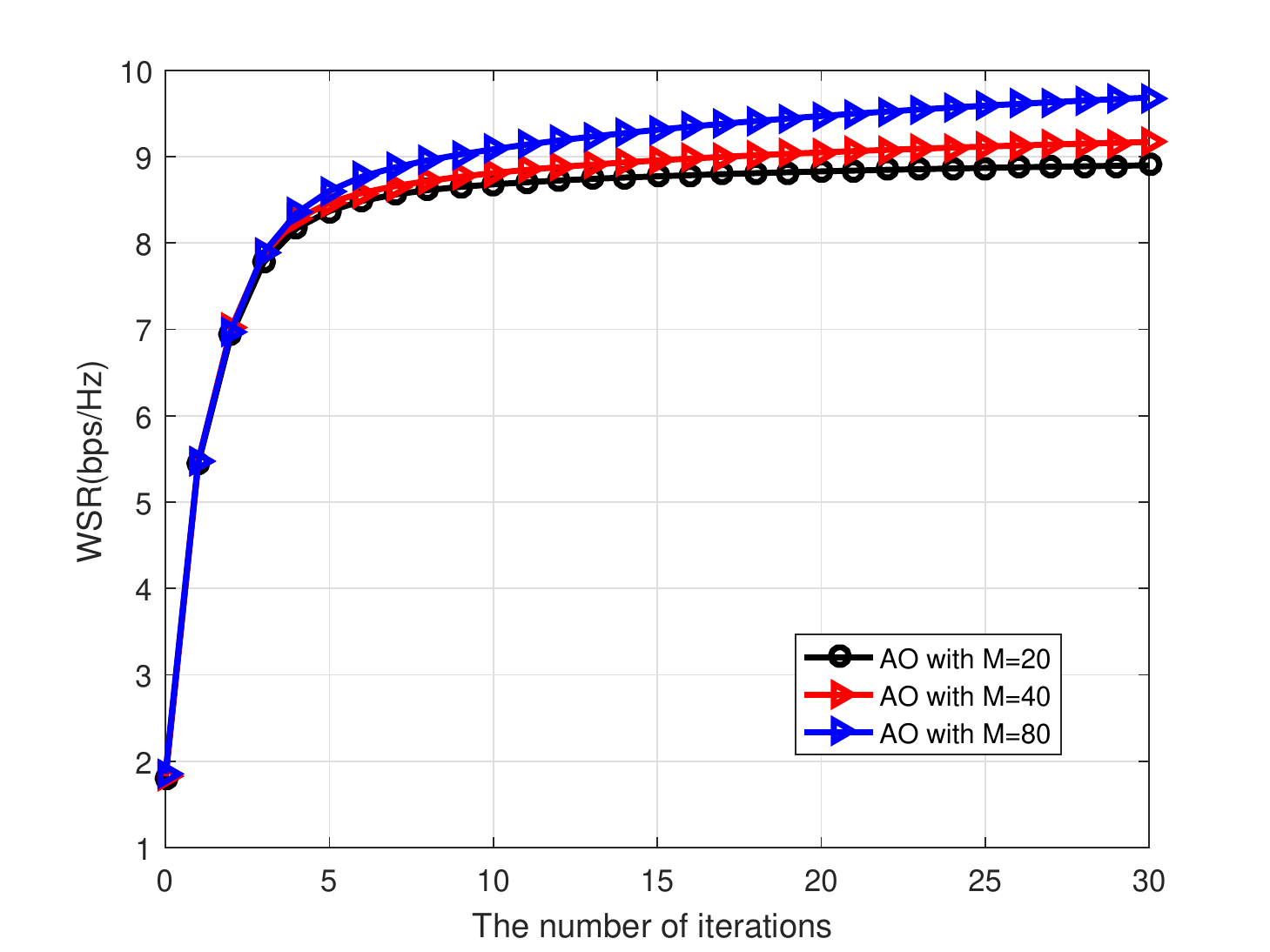}
    \caption{\quad WSR vs the number of iterations}
    \label{RateVsItera_M}
\end{figure}

In order to evaluate the convergence of the proposed algorithm, the variation of WSR with the number of iterations is given in Fig. \ref{RateVsItera_M}. Considering algorithms `AO with $M=20$', `AO with $M=40$' and `AO with $M=80$', the performance corresponding to 30 iterations is calculated in the figure. Note that, the proposed algorithm can converge quickly, i.e., no more than 10 outer iterations can surely promise the convergence of the AO algorithm. In addition, with the number of IRS reflection elements increases, the performance of the proposed algorithm is also improved, which further proves that the wireless environment can be improved by appropriately increasing the number of IRS reflection elements.

\subsection{Special Scenario With Only One PU}
In this subsection, the simulation results about the special scenario are given. Note that, the location of the unique PU is set as $(50, 0)$, whereas the other system parameters are the same as those in the general scenario. Furthermore, to prove the performance gain of the proposed algorithms, two benchmark algorithms used in last subsection are retained but the algorithm `AO' is used to indicate the proposed Algorithm 9. Meanwhile, with the same evaluated parameters, the simulation results shown below are the same as those obtained in the last section, which means that Algorithm 9 has a much lower computational complexity than Algorithm 3 without the performance loss.

\begin{figure}[h]
    \centering
    \includegraphics[scale=0.6]{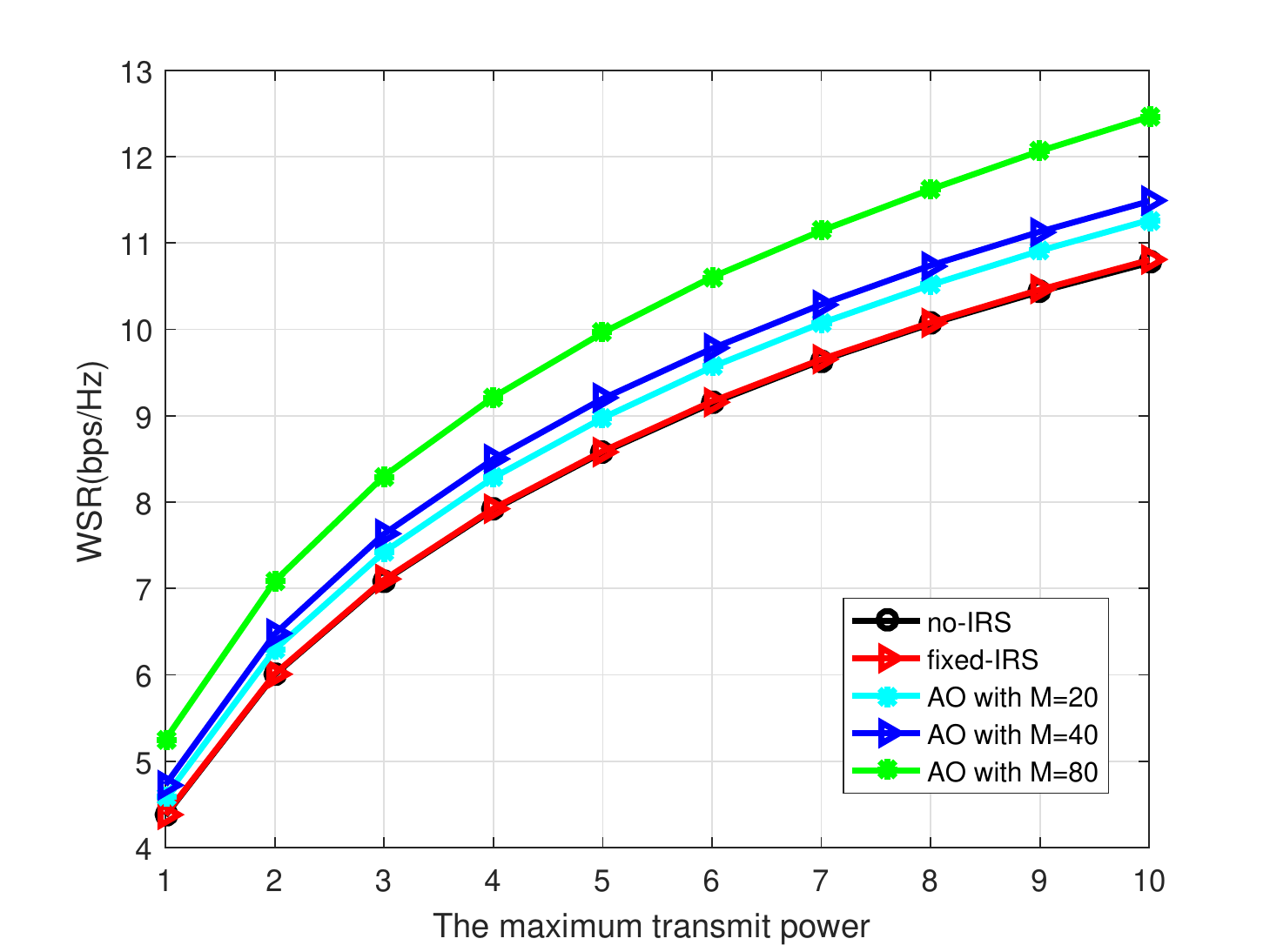}
    \caption{\quad WSR vs the maximum transmit power}
    \label{RateVsPower_S}
\end{figure}
Similarly, the WSR of different algorithms is evaluated by varying the available transmission power at the SAP, i.e., the $P_{max} \in [1,10]W$, and the result is shown in Fig. \ref{RateVsPower_S}. The `AO with $M = 20$', `AO with $M = 40$', and `AO with $M = 80$' shown in the figure are used to identify Algorithm 9 when the number of IRS reflection elements is 20, 40, and 80, respectively. Clearly, the WSR for all these algorithms increases with the increase of the available transmission power at the SAP. Moreover, algorithms `no-IRS' and `fixed-IRS' obtain the worst performance and they are exceedingly close. Introducing the IRS optimization, the performance of `AO' is improved dramatically with the increase of the number of IRS reflection elements and is better than that of `no-IRS' and `fixed-IRS'. It is worth mentioning that the performance gaps between the various mechanisms become larger with the increase of the transmission power at the SAP, which means that the introduction of the IRS reflection coefficient optimization brings more significant performance gains at the higher transmission power.

\begin{figure}[h]
    \centering
    \includegraphics[scale=0.6]{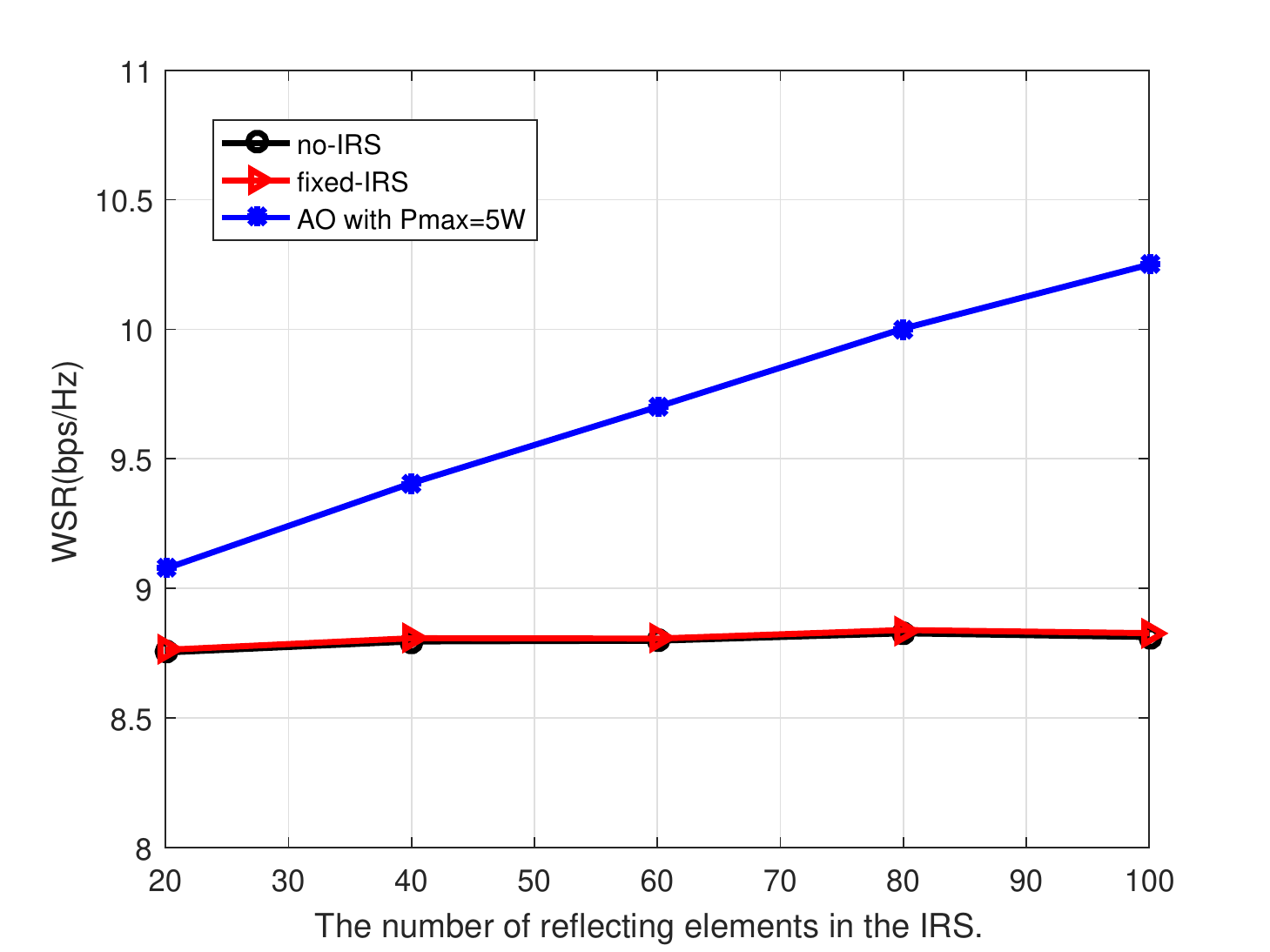}
    \caption{\quad WSR vs the number of reflecting elements}
    \label{RateVsIRS_S}
\end{figure}
Then, Fig. \ref{RateVsIRS_S} shows the trend of the WSR with the number of IRS reflection elements under three mechanisms to prove the impact of the number of IRS reflection elements visually. The maximum transmission power of the SAP is set as $P_{\max} = 5W$, and the number of IRS reflection elements varies from 20 to 100. The `AO with $P_{\max} = 5W$' shown in the figure is used to identify algorithm 9 with $P_{\max} = 5W$. Note that, compared to other two mechanisms, the performance of `AO with $P_{\max} = 5W$' is significantly improved, while the performance of `no-IRS' and `fixed-IRS' mechanisms is not affected by the number of the IRS elements in the system. The fact means that the wireless communication environment can be improved by increasing the number of IRS reflection elements appropriately.

\begin{figure}[h]
    \centering
    \includegraphics[scale=0.6]{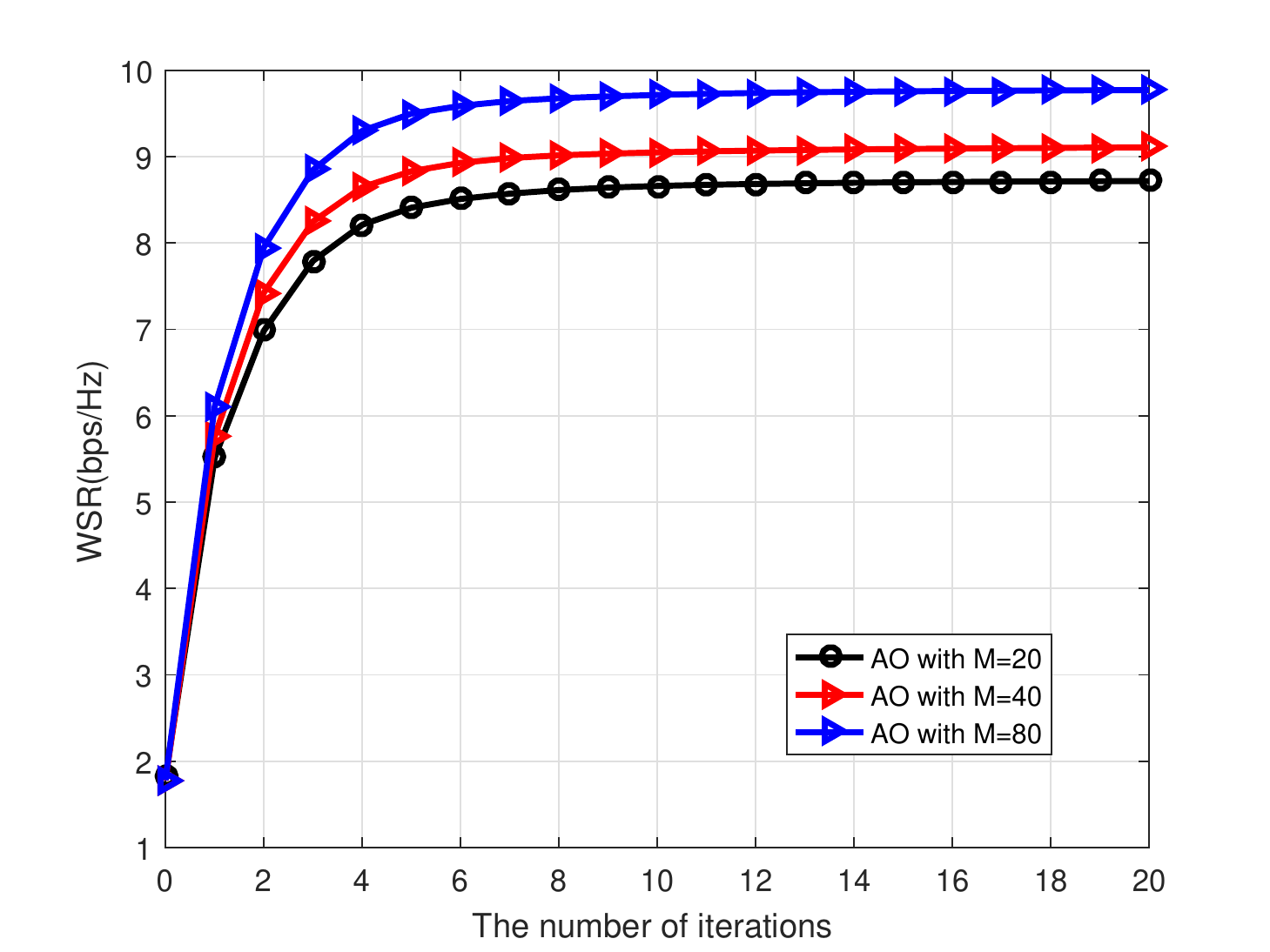}
    \caption{\quad WSR vs the number of iterations}
    \label{RateVsItera_S}
\end{figure}
Finally, the convergence behavior of the proposed algorithm 9 is evaluated and the result is shown in Fig. \ref{RateVsItera_S}. Considering algorithms `AO with $M=20$', `AO with $M=40$' and `AO with $M=80$', the performance corresponding to 30 iterations is calculated in the figure. Note that, the performance of the proposed algorithm is improved slowly after more than 6 iterations, i.e., the proposed algorithm can converge quickly. In addition, with the number of IRS reflection elements increases, the performance of the proposed algorithm is also improved, which further proves that the wireless environment can be improved by appropriately increasing the number of IRS reflection elements. It is worth mentioning that Algorithm 9 has a faster convergence speed than Algorithm 3.

\section{Conclusion}
In this paper, the joint transmit precoding and reflect beamfroming for the IRS-MIMO-CR system is proposed and analyzed. Our design objective is to maximize the achievable WSR of SUs by jointly optimizing the transmit precoding matrices at the SAP and the reflecting coefficients at the IRS, subject to a total transmit power constraint at the SAP and interference constraints at PUs. Since the formulated problem is non-convex with coupled variables, thus the WMMSE is adopted to transfer it to a tractable one and then an AO-based algorithm is proposed. Furthermore, a special scenario with only one PU is considered and an AO-based algorithm with lower complexity is presented. Numerical simulation results confirm that the proposed algorithms can obtain significantly performance gain over the benchmark schemes. In addition, for our considered scenario, the beamforming optimization at IRS can bring much more performance improvement as higher transmission power is allowed at the SAP. It is important to note that, the work in this paper is based on the hypothesis of perfect CSI. However, the channel estimation is bound to have a certain degree of error in practice. Therefore, the problem with imperfect CSI is more interesting and this is one of our future work.

\appendices
\section{Proof of Proposition 1}\label{appendix_1}
Proof: Herein, let $\mathbf{T} = \mathbf{Z}_p - \mathbf{X}_p$ and $f\left( \mathbf{F} \right) = \sum\limits_{l = 1}^L { Tr \left( \mathbf{F}_l^H \mathbf{T} \mathbf{F}_l \right)}$. Given $\mathbf{F}^{(n)}$, the first order Taylor expansion of the function $f\left( \mathbf{F} \right)$ is
\begin{equation}\nonumber
\tilde f\left( \mathbf{F} | \mathbf{F}^{(n)} \right) \buildrel \Delta \over = \sum\limits_{l = 1}^L{ 2 Re \left\{ {Tr \left( \mathbf{F}_l^{(n)H} \mathbf{T} \mathbf{F}_l \right)} \right\}} - \sum\limits_{l = 1}^L {Tr \left( \mathbf{F}_l^{(n)H} \mathbf{TF}_l^{(n)} \right)}.
\end{equation}
Hence, we have ${\tilde f}\left( {{\mathbf{F}^{(n)}}|{\mathbf{F}^{(n)}}} \right) = f \left( {{\mathbf{F}^{(n)}}} \right)$ and ${\left. {{\nabla _{{\mathbf{F}^*}}}{\tilde f}\left( {\mathbf{F}|{\mathbf{F}^{(n)}}} \right)} \right|_{\mathbf{F} = {\mathbf{F}^{(n)}}}} = {\left. {{\nabla _{{\mathbf{F}^*}}}f\left( \mathbf{F} \right)} \right|_{\mathbf{F} = {\mathbf{F}^{(n)}}}}$. Substitute $\mathbf{T} = \mathbf{Z}_p - \mathbf{X}_p$ into the previous equation, the conditions 1) and 2) are proved.

Moreover, since $\mathbf{Z}_p = {\lambda}_p \mathbf{I}$ and ${\lambda}_p$ is the maximum eigenvalue of the $\mathbf{X}_p$, $\mathbf{T} \succcurlyeq \mathbf{0}$. which means that $f \left( \mathbf{F} \right)$ is the convex function with respect to $\mathbf{F}$. Hence,
$$f \left( \mathbf{F} \right) \ge \tilde f \left( {\mathbf{F}|{\mathbf{F}^{(n)}}} \right) \;.$$

Similarly, substituting $\mathbf{T}$ into the above equation and transfer the term, we obtain the condition 3). That is, we have the proposition.  $\hfill\blacksquare$

\section{Proof of Proposition 2}\label{appendix_2}
Proof: Consider two dual variables $\mu _1$ and $\mu _2$ where $\mu _2 > \mu _2$. Let $\mathbf{\hat F}\left( \mu _1 \right)$ and $\mathbf{\hat F}\left( \mu _2 \right)$ be the optimal solutions of problem (\ref{eq_scene1_dual_func}) with $\mu _1$ and $\mu _2$, respectively. Since $\mathbf{\hat F}\left( \mu _1 \right)$ is the optimal solution of (\ref{eq_scene1_dual_func}) with $\mu = \mu _1$, we have
$$L\left( {\mathbf{\hat F}\left( \mu _1 \right),{\mu _1}} \right) \le L\left( \mathbf{\hat F}\left( \mu _2 \right), \mu _1 \right) \;.$$

Meanwhile, we also have
$$L\left( {\mathbf{\hat F}\left( \mu _2 \right),{\mu _2}} \right) \le L\left( \mathbf{\hat F}\left( \mu _1 \right), \mu _2 \right) \;.$$
By adding these two inequalities and simplifying them, we have $\left( \mu _1 - \mu _2 \right) J\left( \mu _1 \right) \le \left( \mu _1 - \mu _2 \right) J\left( \mu _2 \right)$. Since $\mu _1 > \mu _2$, we have $J\left( \mu _1 \right) \le J\left( \mu _2 \right)$. Therefore, we have this proposition. $\hfill\blacksquare$

\section{Proof of Proposition 4}\label{appendix_3}
Proof: Note that, Algorithm 6 obtains the optimal solution of (\ref{eq_appro_pro}) in each iteration.
Based on the fact, denote the convergent solution obtained by Algorithm 6 as $\mathbf{\hat F}$. Moreover, let $f_0\left( \mathbf{F} \right)$, $f_1\left( \mathbf{F} \right)$ and $f_2\left( \mathbf{F} \right)$ denote the objective function, left parts of the power constraint and the interference power constraint of the OP7, respectively.

Given the initial point $\mathbf{\hat F}$, we construct the problem (\ref{eq_appro_pro}) and let ${\tilde f_2}\left( {\mathbf{F}|\mathbf{\hat F}} \right)$ denote the left part of the approximate interference power constraint. The optimal solution of the problem is $\mathbf{\hat F}$ since it is the convergent solution obtained by Algorithm 6. Now, there exist ${\lambda ^*} \ge 0$ and ${\mu ^*} \ge 0$ which satisfy the KKT conditions of the problem (31) as follows
$$\left\{ \begin{array}{l}
 {f_1}\left( {\mathbf{\hat F}} \right) \le {P_{\max }} \\
 {{\tilde f}_2}\left( {\mathbf{\hat F}|\mathbf{\hat F}} \right) \le {{\tilde \Gamma }_p} \\
 {\lambda ^*}\left( {{f_1}\left( {\mathbf{\hat F}} \right) - {P_{\max }}} \right) = 0 \\
 {\mu ^*}\left( {{{\tilde f}_2}\left( {\mathbf{\hat F}|\mathbf{\hat F}} \right) - {{\tilde \Gamma }_p}} \right) = 0 \\
 \nabla {f_0}\left( {\mathbf{\hat F}} \right) + {\lambda ^*}\nabla {f_1}\left( {\mathbf{\hat F}} \right) + {\mu ^*}\nabla {{\tilde f}_2}\left( {\mathbf{\hat F}|\mathbf{\hat F}} \right) = 0 \\
 \end{array} \right. \;.$$

In addition, given $\mathbf{\hat F}$, there are ${f_2}\left( {\mathbf{\hat F}} \right) = {\tilde f_2}\left( {\mathbf{\hat F}|\mathbf{\hat F}} \right) + \sum\limits_{l = 1}^L { Tr \left( {\mathbf{F}_l^{(n)H}\left( {{\mathbf{Z}_p} - {\mathbf{X}_p}} \right)\mathbf{F}_l^{(n)}} \right)}$ and $\nabla {f_2}\left( {\mathbf{\hat F}} \right) = \nabla {\tilde f_2}\left( {\mathbf{\hat F}|\mathbf{\hat F}} \right)$, so the KKT conditions of the OP7 is satisfied with $\lambda ^*$ and $\mu ^*$ as follows,
 $$\left\{ \begin{array}{l}
 {f_1}\left( {\mathbf{\hat F}} \right) \le {P_{\max }} \\
 {f_2}\left( {\mathbf{\hat F}} \right) \le {\Gamma _p} \\
 {\lambda ^*}\left( {{f_1}\left( {\mathbf{\hat F}} \right) - {P_{\max }}} \right) = 0 \\
 {\mu ^*}\left( {{f_2}\left( {\mathbf{\hat F}} \right) - {\Gamma _p}} \right) = 0 \\
 \nabla {f_0}\left( {\mathbf{\hat F}} \right) + {\lambda ^*}\nabla {f_1}\left( {\mathbf{\hat F}} \right) + {\mu ^*}\nabla {f_2}\left( {\mathbf{\hat F}} \right) = 0 \\
 \end{array} \right. \;.$$

Meanwhile, the OP7 is a convex optimization problem. Hence, $\mathbf{\hat F}$ obtained by algorithm 6 is the optimal solution of the OP7 \cite{DBLP:books/cu/BV2014}. Therefore, we have this proposition.  $\hfill\blacksquare$

\ifCLASSOPTIONcaptionsoff
  \newpage
\fi

\bibliographystyle{IEEEtran}
\bibliography{bibfile}

\end{document}